\documentclass[fleqn,usenatbib]{mnras}

\usepackage[T1]{fontenc}
\usepackage{ae,aecompl}

\usepackage{graphicx}	
\usepackage{amsmath}	
\usepackage{amssymb}	
\usepackage{xcolor}
\usepackage{mathtools}
\usepackage{graphbox}
\usepackage{accents}
\usepackage{siunitx}

\definecolor{deeppurple}{rgb}{0.5,0,1}

\usepackage{newtxtext,newtxmath}

\DeclareSIUnit\angstrom{\text {Å}}

\title[Constraining escape fraction evolution]
{Dark-ages reionization and galaxy formation simulation — XXI\@. Constraining
the evolution of the ionizing escape fraction}

\author[S. J. Mutch et al.]{\
Simon J. Mutch,$^{1,2,3}$\thanks{E-mail: smutch@unimelb.edu.au (SJM)}
Bradley Greig,$^{1,2}$
Yuxiang Qin,$^{1,2}$
Gregory B. Poole,$^{4,1}$
J. Stuart B. Wyithe$^{1,2}$
\\
$^{1}$School of Physics, The University of Melbourne, Parkville, VIC 3010,
Australia\\
$^{2}$ARC Centre of Excellence for All Sky Astrophysics in 3 Dimensions (ASTRO 3D)\\
$^{3}$Melbourne Data Analytics Platform, The University of Melbourne, Parkville, VIC 3010, Australia\\
$^{4}$Centre for Astrophysics and Supercomputing, Swinburne University of
Technology, PO Box 218, Hawthorn, VIC 3122, Australia
}

\date{Accepted XXX\@. Received YYY;\@ in original form ZZZ}

\pubyear{2018}

\begin{document}%
\label{firstpage}
\pagerange{\pageref{firstpage}--\pageref{lastpage}}
\maketitle

\begin{abstract}
The fraction of ionizing photons that escape their host galaxies to ionize hydrogen in the inter-galactic medium (IGM) is a critical parameter in analyses of the reionization era. In this paper we use the \textsc{Meraxes} semi-analytic galaxy formation model to infer the mean ionizing photon escape fraction and its dependence on galaxy properties through joint modelling of the observed high redshift galaxy population and existing constraints on the reionization history. Using a Bayesian framework, and under the assumption that escape fraction is primarily related to halo mass, we find that the joint constraints of the UV luminosity function, CMB optical depth, and the Ly$\alpha$ forest require an escape fraction of $(18\pm5)\%$ for galaxies within haloes of $M\lesssim10^{9}$M$_\odot$ and $(5\pm2)\%$ for more massive haloes. In terms of galaxy properties, this transition in escape fraction occurs at stellar masses of $M_\star\sim10^7$M$_\odot$, nearly independent of redshift. As a function of redshift, reionization is dominated by the smaller $M_\star\lesssim10^7$M$_\odot$ galaxies with high escape fractions at $z\gtrsim6$ and by the larger $M_\star\gtrsim10^7$M$_\odot$ galaxies with lower escape fractions at $z\lesssim6$. Galaxies with star formation rates of $10^{-2.5}$M$_\odot$yr$^{-1}$ to $10^{-1.5}$M$_\odot$yr$^{-1}$ provide the dominant source of ionizing photons throughout reionization. Our results are consistent with recent direct measurements of a $\sim5\%$ escape fraction from massive galaxies at the end of reionization and support the picture of low mass galaxies being the dominant sources of ionizing photons during reionization.
\end{abstract}




\section{Introduction}%
\label{sec:intro}

The Epoch of Reionization, during which the progenitors of modern day galaxies formed and reionized hydrogen throughout the intergalactic medium (IGM), is the least explored cosmic epoch. However, over the past two decades there has been significant progress in exploring the stellar growth history of these galaxies \citep[e.g.][]{Bouwens:2017,Finkelstein:2015}, with a recent step-change in understanding that has arisen through the James-Webb Space Telescope \citep[e.g.][]{NaiduJWST,Atek2022RevealingSMACS0723}. At the same time, observations of the evolution of the ionization state of the inter-galactic medium has been narrowed towards a scenario where reionization completes at, or just below, $z\sim6$ \citep[e.g.][]{Bosman:2022,Qin:2021}. This constraint has been achieved through a combination of the integral measure of the reionization history obtained from CMB measurements \citep[e.g.][]{Planck-Collaboration:2016}, and measurements of absorption signals against sources emitting Ly-$\alpha$ photons, including Ly-$\alpha$ forest optical depth, dark gaps and quasar Ly-$\alpha$ emission line damping wings \citep[e.g.][]{Becker:2013,McGreer:2015,Banados2017,Greig2022}.

Star forming galaxies are thought to provide the dominant source of ionizing photons to reionize the Universe \citep[e.g.][]{Srbinovsky:2007,Bouwens:2015,Qin:2017}. However, we are unable to directly relate the observed star forming galaxy population to the evolution in the ionization state of the IGM due to the unknown fraction of ionizing photons which escape from the ISM of galaxies. Given the evolution of galaxy density with redshift, and the expected dependence of star-formation rate with host-halo mass, we expect that there is likely a dependence of escape fraction on these quantities \citep[e.g.][]{Gnedin:2008,Wise:2009}. For example, in a scenario where the dominant photon sources are low-mass haloes, reionization starts at higher redshift \citep[e.g.][]{Finkelstein:2019}. However, if galaxies in more massive haloes are the more dominant sources, then reionization will be delayed and conclude more rapidly \citep[e.g.][]{Mesinger:2016,Naidu:2020}. Moreover, since this escape of radiation is likely through channels of low column density HI and dust, the escape fraction is expected to be direction dependent for individual galaxies and to have a large scatter with galaxy properties \citep[e.g.][]{Cen:2015, Yeh:2023}.

Given the importance of the escape fraction to our understanding of reionization, there have been a number of studies to i) measure the escape fraction directly in individual galaxies \citep[e.g.][]{Pahl:2022,Begley:2022,Bassett:2022}, ii) to directly simulate the ISM of star forming galaxies in a cosmological context \citep[e.g.][]{Yeh:2023,Wise:2009}, and iii) to infer the average escape fraction properties by comparing observed star-formation rates with the reionization history \citep[e.g.][]{Wyithe:2010,Finkelstein:2019,Yung:2020}.  

In this paper we follow the third approach, using a semi-analytic model coupled with semi-numerical calculations of reionization, which offers a computationally efficient route to calculating the connection between galaxy formation and reionization \citep[e.g.][]{Mutch:2016,Seiler:2019,Visbal:2020,Hutter:2021}. We utilize the \textsc{Meraxes} semi-analytic model \citep{Mutch:2016} to provide a realistic description of the galaxy population at $z\sim4-8$, which can be extended to both redshifts higher than observed and to luminosities fainter than observed based on physical descriptions of galaxy formation physics. \textsc{Meraxes} couples this galaxy evolution to the evolution of the ionization state of the IGM with an assumed escape fraction. Our approach is to provide a flexible relationship between galaxy properties and the escape fraction, and then to constrain the parameters describing this relationship via a Bayesian framework. Efficient implementation of the \textsc{Meraxes} semi-analytic model makes this parameter space exploration possible, even though the model includes on the fly calculation of reionization and is directly run on a dark-matter halo population close to the hydrogen cooling limit within a large N-body simulation.

The paper outline is as follows. In Section~\ref{sec:meraxes} we outline the relevant features of \textsc{Meraxes}. We then describe our implementation of the statistical sampling of the parameter space for \textsc{Meraxes} in Section~\ref{sec:sampling} before constraining our galaxy formation model parameters in Section~\ref{sub:UVLF}. In Section~\ref{sec:constraints} we describe the reionization observations we employ before presenting our escape fraction model and the resulting constraints in Section~\ref{sec:results}. We finally provide our discussion and conclusions in Sections~\ref{sec:discussion} and \ref{sec:conclusions}, respectively. Throughout this paper, we adopt a standard $\Lambda$CDM cosmology with $\Omega_{\rm m} = 0.308$, $\Omega_{\Lambda}=0.692$ and $H_0 = 67.8\,$km/s$\,$Mpc$^{-1}$.

\section{Meraxes}%
\label{sec:meraxes}

 The \textsc{Meraxes} semi-analytic model \citep{Mutch:2016} is designed to explore galaxy formation across cosmic time, with a particular focus on the Epoch of Reionization and the first galaxies.
Some of its key features include a delayed supernova feedback scheme which accounts for the finite lifetime of massive stars and the use of ``horizontal'' merger trees whereby the entire simulation volume is processed one snapshot at a time (necessary for an efficient treatment of reionization).
However, the defining feature of \textsc{Meraxes} is its close coupling to a modified version of the semi-numerical reionization code, \textsc{21cmFAST} \citep{Mesinger:2011}.
This allows coupled modelling of reionization with the growth of the galaxy population, both temporally and spatially.
Taking into account both local and external ionizing radiation, we can explore how subsequent star formation is affected and how this feeds back on to the progression of reionization.

As with many other semi-analytic models, \textsc{Meraxes} includes treatments for star formation, supernova feedback, metal enrichment, cooling, AGN feedback, and galaxy mergers.
In this section we provide a broad overview of the model and any physical prescriptions that are directly relevant for this work.
A full description of the model is provided in \citet{Mutch:2016} and the subsequent papers referenced below.

\subsection{Tiamat}%
\label{sub:tiamat}
Semi-analytic galaxy formation models work by post-processing dark matter halo merger trees either extracted from cosmological N-body simulations or created using Extended Press-Schechter (EPS) theory.
In order to fully model reionization, where the spatial distribution of galaxies directly influences their evolution, the former is required.

In this work, we utilise the Tiamat N-body simulation, run as part of the Dark-ages Reionization And Galaxy Observables from Numerical Simulations programme \citep[DRAGONS; ][]{Poole:2016}.
Tiamat consists of $2160^3$ collisionless particles following a Planck 2015 cosmology \citep{Planck-Collaboration:2016a} in a comoving volume of ${(100\, \rm Mpc)}^3$, resulting in a particle mass of ${\rm 3.89 \times 10^6\, M_\odot}$.
This allows us to identify the majority of atomic-cooling mass haloes across all redshifts during reionization within a large enough volume to provide a statistically representative realisation of the large scale process.
Tiamat has been run down to a redshift of $z$=1.8 with 101 snapshots output between $z$=49 and 5, evenly spaced in time with a cadence of ${\sim 11\, \rm Myr}$, and a further 64 snapshots at ${z<5}$, logarithmically spaced in redshift.
This fine snapshot cadence at high redshift allows us to better track the progression of global reionization, which proceeds rapidly during its mid phases.
It also allows us to more accurately model the stochastic nature of star formation and its impacts on the reionization process at these high-redshifts \citep{Mutch:2016}.
For more details on the Tiamat simulation, interested readers are referred to \citet{Poole:2016} and \citet{Angel:2016}.
Details on the halo merger tree construction can be found in \citet{Poole:2017}.

\subsection{Star formation and supernova feedback}%
\label{sub:star_formation}

The star formation and supernova feedback prescriptions implemented in \textsc{Meraxes} closely follow those of \citet{Croton:2006}, with updates presented in \citet{Guo:2011}.

At each time step in the simulation, we calculate the amount of gas required to be accreted by each host dark matter halo in order to maintain the required baryon fraction in the presence of the intergalactic ionization field.
This gas is assumed to be shock-heated to the virial temperature of the halo before cooling down to the central regions to form a rotationally supported exponential disc.
Assuming angular momentum conservation of the gas as it cools, the scale radius of this disc is given by $r_{\rm disc} = R_{\rm vir}\lambda / \sqrt{2}$, where $\lambda$ is the dimensionless spin parameter of the halo \citep{Bullock:2001}.

Following \citet{Kennicutt:1989}, we assume that star formation in the disc only occurs once the cold gas density reaches a critical threshold, $\Sigma_{\rm crit}$, which can be linked to the instantaneous properties of the host dark matter halo.
The resulting star formation rate, $\dot m_*$, is calculated by assuming that some fraction, $\alpha_{\rm SF}$, of the cold gas, $m_{\rm cold}$, above this threshold is then turned into stars per disc dynamical time, $t_{\rm dyn}$:
\begin{equation}
  \label{eqn:star_formation}
  \dot m_* = \frac{\alpha_{\rm SF}}{t_{\rm dyn}}  {\left(m_{\rm cold} - 2\pi \Sigma_{\rm crit} r_{\rm disc}^2 \right)}\,,
\end{equation}
where the second term on the right-hand side corresponds to the cold gas mass resulting in the critical surface density for a disc with scale radius $r_{\rm disc}$.

Galaxy mergers can also drive shocks and turbulence, resulting in enhanced star formation episodes. We model the fraction of cold gas converted into stars in such events following \cite{Somerville:2001}:
\begin{equation}
  \frac{m_{\rm burst}}{m_{\rm cold}} = \alpha_{\rm burst} {\left(\frac{m_{\rm gal}}{m_{\rm parent}}\right)}^{\gamma_{\rm burst}}\ ,
\end{equation}
where $m_{\rm burst}$ is the mass of newly formed stars, $m_{\rm gal}$ and $m_{\rm parent}$ are the total baryonic masses of the infalling and parent galaxies respectively, and $\gamma_{\rm burst}$ and $\alpha_{\rm burst}$ are free parameters.
Following \cite{Croton:2006}, we fix ${\gamma_{\rm burst}=0.7}$ and ${\alpha_{\rm burst}=0.57}$.

Energy injection from {supernov\ae} is thought to be the dominant feedback mechanism for controlling the growth of stellar mass (and hence the production rate of ionizing photons) at the high redshifts and low halo masses relevant for reionization \citep[see e.g.][]{Qin:2017}.
We assume that this feedback mechanism is dominated by massive, short-lived (${\lesssim 40\, {\rm Myr}}$) stars, ending in energetic Type-II {supernov\ae}.
The total supernova energy injected into the ISM, ${\Delta E_{\rm total}}$, after a time, $t$, from past star formation episodes of mass ${\delta m_*}$ is given by:
\begin{equation}
  \label{eqn:sn_energy_coupling}%
  \Delta E_{\rm total} = \epsilon \int_0^t \frac{\delta m_*}{dt'} d\xi(t') dt'\,,
\end{equation}
where ${d\xi}$ is the metallicity dependent energy injection from stars of age $t-t'$ which we calculate directly from the stellar population synthesis code \textsc{Starburst99} \citep{Starburst99}, assuming a \citet{Kroupa:2001} IMF.\@
$\epsilon$ is a free parameter which controls the efficiency with which the injected energy couples to the surrounding ISM.\@
The amount of gas assumed to be directly heated by this energy injection is modelled as
\begin{equation}
  \label{eqn:sn_mass_loading}%
  \Delta m_{\rm total} = \eta \delta m_*\,,
\end{equation}
where $\eta$ is a free parameter commonly referred to as the mass loading factor.

Following \citet{Qiu:2019}, we assume $\epsilon$ and $\eta$ to take the following functional forms:
\begin{equation}
 \label{eqn:epsilon}
  \epsilon = \epsilon_0 {\left(\frac{1+z}{4}\right)}^{\alpha_{\rm eject}} {\left(\frac{V_{\max}}{60\ {\rm km\,s^{-1}}}\right)}^{\beta}\ {\rm and}
\end{equation}
\begin{equation}
  \label{eqn:eta}
  \eta = \eta_0 {\left(\frac{1+z}{4}\right)}^{\alpha_{\rm reheat}} {\left(\frac{V_{\max}}{60\ {\rm km\,s^{-1}}}\right)}^{\beta}\,,
\end{equation}
where $V_{\rm max}$ is the maximum circular velocity of the host halo, $\beta = -3.2\ (-1)$ for $V_{\max} < (\geq)\ 60\ {\rm km\,s^{-1}}$ \citep[following][]{Muratov:2015} and ${\alpha_{\rm reheat}=2}$ \citep[following][]{Cora:2018}.
We also assume ${\alpha_{\rm eject}=0}$, as suggested by \citet{Qiu:2019}.

The energy injected into the ISM is used to adiabatically heat the gas in the galaxy to the virial temperature of the host dark matter halo where it joins a hot gas reservoir.
The mass of gas added to the hot reservoir, ${\Delta m_{\rm hot}}$, is given by
\begin{equation}
  \Delta m_{\rm hot} = \frac{2\Delta E_{\rm total}}{V_{\rm vir}^2} \Delta m_{\rm total}\,.
\end{equation}
If there is more energy available than is required to reheat the full $\Delta m_{\rm total}$ worth of gas to the virial temperature (i.e. ${\Delta m_{\rm hot} > \Delta m_{\rm total}}$) then the remainder of the energy is used to eject the remaining mass from the system entirely.
For more details on the supernova feedback prescription, see \citet{Qiu:2019}.

\subsection{Reionization}%
\label{sub:reionization}

One of the key features of \textsc{Meraxes} is its treatment of reionization, which is modelled using a modified version of the semi-numerical reionization code \textsc{21cmFAST} \citep{Mesinger:2011}.
At each time step in the simulation, Cartesian grids of halo mass, stellar mass, star formation rate, and total matter density are constructed. For this work we employ a grid resolution of $256^3$, corresponding to a side length of 0.39\, Mpc.
These are then filtered using an excursion set formalism to identify ionized cells where the number of ionizing photons is greater than the number of neutral atoms and recombination events.
As discussed in Section~\ref{sec:intro}, the key free parameter controlling this process is the escape fraction of ionizing radiation from the source galaxies, $f_{\rm esc}$.

When a patch of the IGM is reionized by UV photons, the temperature of the gas is raised to $\sim 2 \times 10^4\,{\rm K}$.
This acts to raise the local Jeans mass, reducing the amount of material accreted by low mass haloes.
We model this process using a baryon fraction modifier:
\begin{equation}
  \Delta m_{\rm baryon} = f_{\rm b} f_{\rm mod} M_{\rm vir} - m_{\rm
  baryon}\,,
\end{equation}
where $\Delta m_{\rm baryon}$ is the freshly accreted baryon mass, $f_{\rm b}=\Omega_{\rm b}/\Omega_{\rm m}$ is the universal baryon fraction, $m_{\rm baryon}$ is the mass of baryons already present in the halo, and $f_{\rm mod}$ is the baryon fraction modifier.
The value of $f_{\rm mod}$ is calculated following \citet{Sobacchi:2013} and depends on the mass of the halo, the strength of the local UV ionizing background (which is again strongly dependent on the ionizing escape fraction), and the time over which the halo has been exposed to this background.
For further details of the reionization prescription, see \citet{Mutch:2016}.

\subsection{UV luminosities}

In order to generate UV luminosities for each galaxy at a time $t$, we sum the luminosities from each past star formation burst of age $\tau$. These luminosities are calculated by constructing a full spectral energy distribution (SED) and applying a tophat filter centred on $1600\si{\angstrom}$, with a width of $100\si{\angstrom}$:
\begin{multline}
  L_{\rm UV}(t) = \\
  \sum_{i=0}^{n}\int_{1550\si{\angstrom}}^{1650\si{\angstrom}} d\lambda \int_{0}^{t} dt' \int_{Z_{\rm min}}^{Z_{\rm max}} dZ \psi{\left(t'-\tau_i, Z\right)}S_{\lambda}(\tau_i, Z)T_{\lambda}(\tau_i)\,,
\end{multline}
where $\psi{\left(t'-\tau_i, Z\right)}$ is the metallicity ($Z$) dependant star formation rate of a burst ($i$) of age $\tau_i$, $S_{\lambda}(\tau_i, Z)$ is the wavelength and metallicity dependent luminosity of a simple stellar population of age $\tau_i$, and $T_{\lambda}(\tau_i)$ is the dust transmission function.
We calculate $S_{\lambda}(\tau_i, Z)$ using the publicly available \textsc{starburst99} stellar population synthesis model \citep{Starburst99}, including nebular emission lines and assuming a \citet{Kroupa:2001} IMF with ${Z_{\rm min}=0.001}$ and ${Z_{\rm max}=0.04}$.

\citet{Qiu:2019} explored a number of physically motivated parametrisations for the dust transmission, $T_{\lambda}(\tau)$, by constraining \textsc{Meraxes} to match both the evolution of the observed ultraviolet luminosity function (UVLF) and infrared excess (IRX)-$\beta$ relation at ${z \gtrsim 4}$.
For this work, we employ the gas column density dust model whereby the optical depth, $\Gamma_\lambda$, is given by
\begin{equation}
  \label{eqn:optical_depth}%
  \Gamma_{\lambda} = \mathrm{e}^{-a z}\left(\frac{m_{\mathrm{cold}}}{10^{10} h^{-1} \mathrm{M}_{\odot}}\right)^{\gamma_{\rm GCD}}\left(\frac{r_{\mathrm{disc}}}{h^{-1} \mathrm{kpc}}\right)^{-2} \left(\frac{\lambda}{1600 \si{\angstrom}}\right)^{n}\,,
\end{equation}
where $a$, $\gamma_{\rm GCD}$ and $n$ are free parameters and $r_{\rm disc}$ and $m_{\rm cold}$ are as introduced above.
The resulting transmission is calculated using the two phase model of \citet{Charlot:2000}:
\begin{equation}
  \label{eqn:transmission}%
  T_{\lambda}(t) =
  \begin{cases}
    \exp \left(-\left(\tau_{\mathrm{ISM}}-\tau_{\mathrm{BC}}\right) \Gamma_\lambda\right) & t<t_{\mathrm{BC}} \\
    \exp \left( -\tau_{\mathrm{ISM}}\Gamma_\lambda \right) & t \geq t_{\mathrm{BC}}
  \end{cases}
  \,,
\end{equation}
where $\tau_{\rm ISM}$ and $\tau_{\rm BC}$ are again free parameters.
Here, photons emitted by young stars with ages less than $t_{\rm BC}$ must travel through both a dense birthcloud and the more extended ISM.
After this time, the birthcloud is assumed to have been destroyed and the photons are only attenuated by the ISM.
For more detail on both the SED construction and dust model see \citet{Qiu:2019}.

\subsection{Unique features for parameter space exploration}

One of the key benefits of semi-analytic models over more detailed hydrodynamic simulations is their relatively low computational expense.
This opens up the potential for statistical exploration of the underlying free parameter space \citep[e.g.][]{Henriques:2013,Qiu:2019}.
\textsc{Meraxes} has been designed to maximise this potential by being written to be as efficient as possible without loss of flexibility.
Some of the relevant unique features of our model include the pre-loading of input data, the utilisation of graphics processing units (GPUs), and the suite of companion tools which have been developed to programmaticaly interact with the model and its output.

In order to provide a converged sampling of the underlying posterior probability distribution we require many tens-of-thousands of model evaluations for moderate dimensionality problems ($\sim$10 free parameters).
In this situation, it is imperative that the model being evaluated can be run as quickly and efficiently as possible.
For \textsc{Meraxes} (running on our input N-body simulation, Tiamat) at least $\sim$70\% of the time taken for a single model call is devoted to reading in the trees and dark matter density grids from disk, distributing them amongst all available processors, and processing their contents.
Therefore, we have modified the SAM to be able to read in all input data once, store this in memory, and then to run multiple times without the need for any expensive input-output (IO) calls.

A further computational bottleneck for \textsc{Meraxes} is its treatment of reionization using a modified version of the 21cmFast \citep{Mesinger:2011} semi-numerical code.
This currently requires the model to filter a number of 3D Cartesian grids using an excursion set formalism at every snapshot.
Each of these excursion set calculations entails a large number of Fourier transforms and multiple loops over every cell (of which there are $256^3$ for this work) in the gridded volume.
To reduce the amount of time spent doing this expensive calculation, and thus allow our analysis to be completed in a timely fashion, the reionization portion of \textsc{Meraxes} has been re-written to take advantage of the Graphics Processing Units (GPUs) commonly available on the majority of modern high performance computing systems.
This has sped up the reionization calculation by a factor of 8 for typical combinations of grid dimensions and hardware resources.

\section{Parameter space sampling}%
\label{sec:sampling}

In order to carry out a statistical exploration of the \textsc{Meraxes} parameter space, we require a method to calculate the likelihood of the model predictions for a given set of parameters and to then provide the code with a new set of trial parameters for the next run.
To facilitate this, as well as future parameter space exploration and calibration studies with \textsc{Meraxes}, we have developed a new companion code called \textsc{Mhysa} which provides a flexible two-way communication layer between the model (written in \textsc{C}) and Python.
\textsc{Mhysa} was designed to efficiently facilitate calling \textsc{Meraxes} many times, with different parameter values, and allows us to interface any Python optimisation or sampling package with the model.
It also allows us to write all of our likelihood calculations in Python and to leverage its many available scientific and numerical libraries, vastly reducing development time.
In practice, \textsc{Mhysa} deals with initialising the model, updating its parameters, directly accessing in-memory galaxies and haloes, calling arbitrary Python code at the end of each snapshot, doing calculations on an independent processor in the background whilst the model is running, and orchestrating multiple instances of the model running concurrently.

A further key feature of \textsc{Mhysa} which greatly enhances future science investigations with \textsc{Meraxes}, is that it is not only limited to statistical sampling.
Often we wish to simply find the `best'' model which matches our chosen observational constraints, without necessarily investigating parameter degeneracies or uncertainties.
\textsc{Mhysa} has been written with a flexible interface that can be used to carry out such a global optimisation of the model parameters using any suitable technique, and leveraging the many publicly available packages.

For this work, we make use of the publicly available Python package \textsc{Ultranest} \citep{Buchner:2021} to carry out our parameter space sampling using the nested sampling \citep{Skilling:2004} Monte Carlo algorithm \textsc{MLFriends} \citep{Buchner:2014,Buchner:2017}.
The benefit of nested sampling over more traditional Markov Chain Monte Carlo (MCMC) methods for this work is four-fold:
Nested sampling does not require us to have an estimate of the global maximum likelihood parameter values in order to avoid a lengthy burn-in phase;
it can more efficiently deal with multi modality or complicated degeneracies in probability distributions;
it typically requires fewer likelihood evaluations to probe the underlying posterior probability distribution to a satisfactory degree of accuracy;
and it naturally provides us with an estimate of the Bayes evidence that can be used to quantitatively compare the support for different escape fraction parametrisations.

For each of the parameter space sampling runs presented in this work, we use a minimum of 200 live points. This is the number of points which are maintained throughout the sampling process and iteratively updated to sample the posterior. More live points result in a higher resolution sampling of the posterior, at the expense of more likelihood evaluations.
The default \textsc{Ultranest} stopping criteria were employed, with the exception of the maximum fraction of the remaining evidence and minimum effective sample size, which we set to 0.1 and 600, respectively.

\section{The Galaxy UV luminosity function and $\beta-M_{\rm UV}$ relation}%
\label{sub:UVLF}

The galaxy UV luminosity function is the most direct observational probe of the reionizing galaxy population at ${z \geq }5$.
It traces the recent star formation at a fixed number density and hence, the total ionizing photon production rate and its spatial distribution \citep[under the assumption that stars provide the bulk of ionizing photons during the EoR, e.g.][]{Qin:2017}. 
Therefore, if we are to carry out a quantitative investigation of allowed escape fraction models, we must first calibrate our model against the observed UVLF.

\begin{table}
 \caption{The free galaxy formation and dust model parameters used in this work when constraining against the observed UVLF and $\beta - M_{\rm UV}$ relations. See \S\ref{sub:UVLF} for details.}%
    \label{tab:model-params}%
    \begin{tabular}{llc}
\hline
Parameter & Description & Equation \\
\hline
\multicolumn{3}{l}{\textit{galaxy formation}} \\
$\log_{10}(\alpha_{\rm SF})$ & Star formation efficiency & \ref{eqn:star_formation} \\
$\log_{10}(\Sigma_{\rm n})$  & Surface density threshold for star formation & \ref{eqn:star_formation} \\
$\log_{10}(\epsilon_0)$      & Supernova energy coupling efficiency & \ref{eqn:epsilon} \\
$\log_{10}(\eta_0)$          & Mass loading factor & \ref{eqn:eta} \\
\\
\multicolumn{3}{l}{\textit{dust}} \\
$\tau_{\rm ISM}$             & ISM optical depth normalisation & \ref{eqn:transmission} \\
$\tau_{\rm BC}$              & Birth cloud optical depth normalistion & \ref{eqn:transmission} \\
$\gamma_{\rm GCD}$           & Cold gas mass dependence & \ref{eqn:optical_depth} \\
$n$                          & Reddening slope & \ref{eqn:optical_depth} \\
$a$                          & Redshift dependence & \ref{eqn:optical_depth} \\
\hline
\end{tabular}
\end{table}

The free parameters of the model which we allow to vary when fitting the UVLF are listed in Table~\ref{tab:model-params}. These are identical to the free parameters used by \citet{Qiu:2019} and include four galaxy formation parameters controlling star formation and supernova feedback, as well as four parameters which control dust extinction and how it varies with cold gas mass. These parameters have been shown in previous works to be the most important for controlling the growth of early galaxies in our model \citep[e.g.][]{Mutch:2016} and reproducing the evolution of the UVLF \citep{Qiu:2019}.

The assumed prior distributions for each parameter are presented in Appendix~\ref{sec:appendix-posteriors}. They are selected to weakly encode our past experience and physical intuition, however, we have ensured that our conclusions are insensitive to our precise choices except where explicitly stated in the text.

For this work, we utilise the observational luminosity functions from \citet{Bouwens:2015}.
Following \citet{Gillet:2020} \citep[and as originally suggested by][]{Bouwens:2017} we enforce a minimum uncertainty of $0.08\,\textrm{dex}$ on all data points.
Since reionization is known to be largely complete to within $x_{\rm HI} \sim 0.05$ by $z \sim 5$ \citep[e.g.][]{Bosman:2022,Qin:2021}, we only consider redshifts greater than this.
The volume of our input N-body simulation ${(100\, \rm Mpc)}^3$ also limits the brightest galaxies for which \textsc{Meraxes} can provide statistically significant predictions, due to the low number density of these biased objects.
Therefore, at each redshift, we only consider luminosities below which observations indicate there should be at least 10 galaxies.

We calculate the likelihood of a model parameter set, $\theta$, assuming uncorrelated, (log-)Normally distributed, errors:
  \begin{multline}
  \label{eq:uvlf-lnL}
    \ln\bigl({\cal L}_{{\rm UVLF}}(\theta)\bigl) = \\
    -\frac{1}{2} \sum_{i=0}^{N_d} \biggl( \frac{{\bigl(\log_{10}(\phi_{i,\rm obs}) - \log_{10}(\phi_{i, {\rm mod}})\bigl)}^2}{\sigma^2_i} - \ln (2 \pi \sigma_i) \biggl)\,,
  \end{multline}
where $\phi_{i,{\rm obs}}$ and $\phi_{i,{\rm mod}}$ are the observations and model predictions for data point $i$, respectively, $\sigma^2_i = \sigma^2_{i,{\rm obs}} + \sigma^2_{i, {\rm mod}}$ is the corresponding variance, and $N_d$ is the total number of data points.
Note that we use the appropriate value of $\sigma_{i,{\rm obs}}$ when the model prediction is greater than or less than the observational data point if the quoted observational uncertainty is non-symmetric. Our choice of uncorrelated log-normal uncertainties is simply our best estimate of the true uncertainty distribution in the absence of further information (such as a covariance matrix) and is a commonly made assumption.
For the model variance, we assume a Poisson distribution with a mean equal to the galaxy number counts in each bin.

In order to fully constrain all eight free model parameters and break degeneracies in the dust extinction model, \citet{Qiu:2019} found it necessary to additionally constrain against the evolution in the UV slope ($\beta$) as a function of magnitude ($M_{\rm UV}$). We do similarly here, utilising the biweight mean observations of \citet{Bouwens:2014} and again assuming uncorrelated, (log-)Normally distributed errors as per Equation~\ref{eq:uvlf-lnL}.

\subsection{Constraints on galaxy formation parameters}%
\label{sub:constraints_on_galform}

\begin{figure}
  \centering
  \includegraphics[width=1.0\columnwidth]{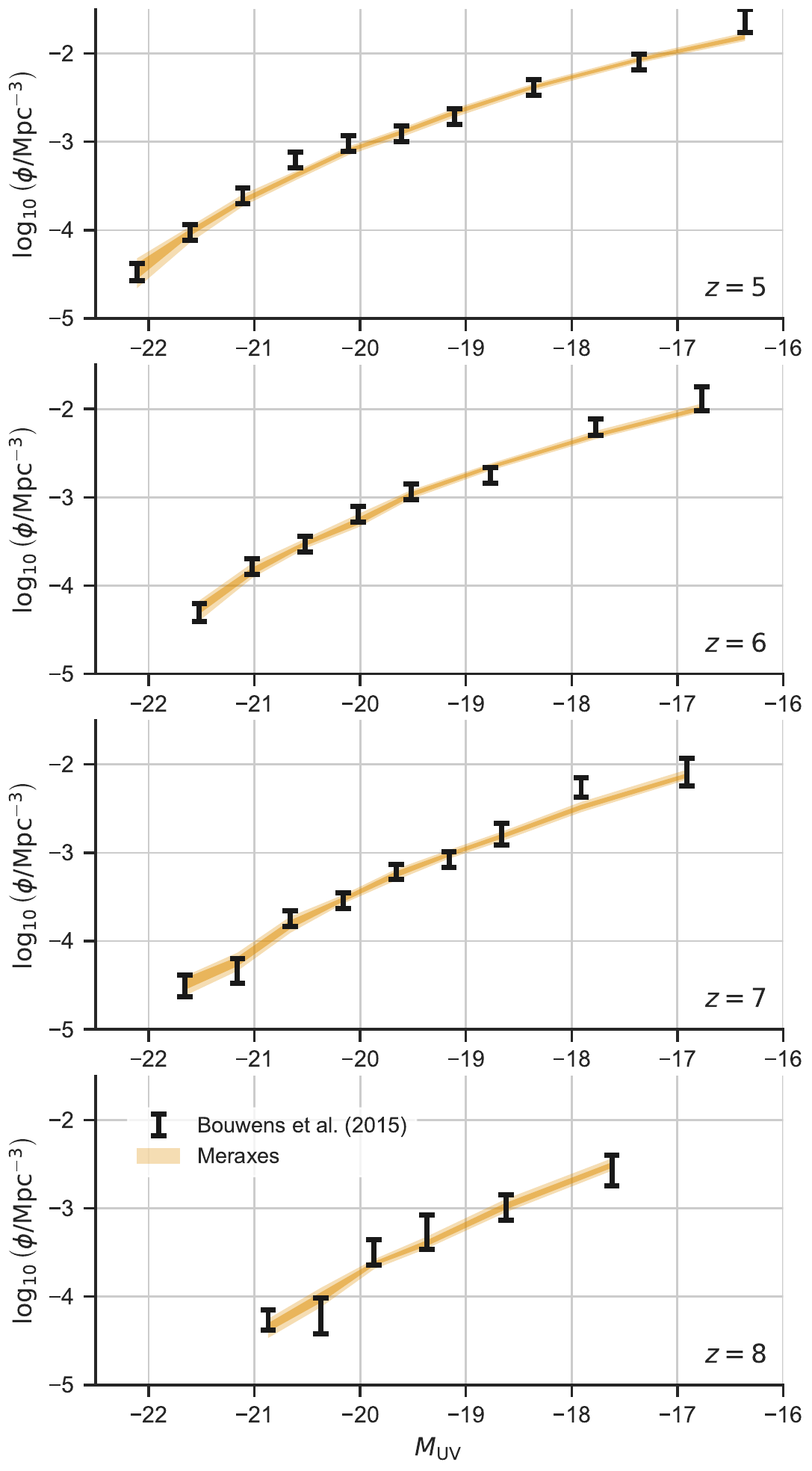}
 \vspace{-5mm}
 \caption{%
    The best-fit galaxy UVLF (shaded regions) obtained by fitting against the observed evolution of the UVLF alone (error bars).
    Dark and light shading indicates the 68 and 95\% confidence intervals, respectively.
    Our model does a reasonable job of matching the observations at all fitted redshifts, providing confidence that we can correctly model the evolution of the observed faction of the total ionizing photon budget.
    See \S\ref{sub:constraints_on_galform} for more details.
  }\label{fig:uvlf-pred}
\end{figure}

\begin{figure}
  \centering
  \includegraphics[width=1.0\columnwidth]{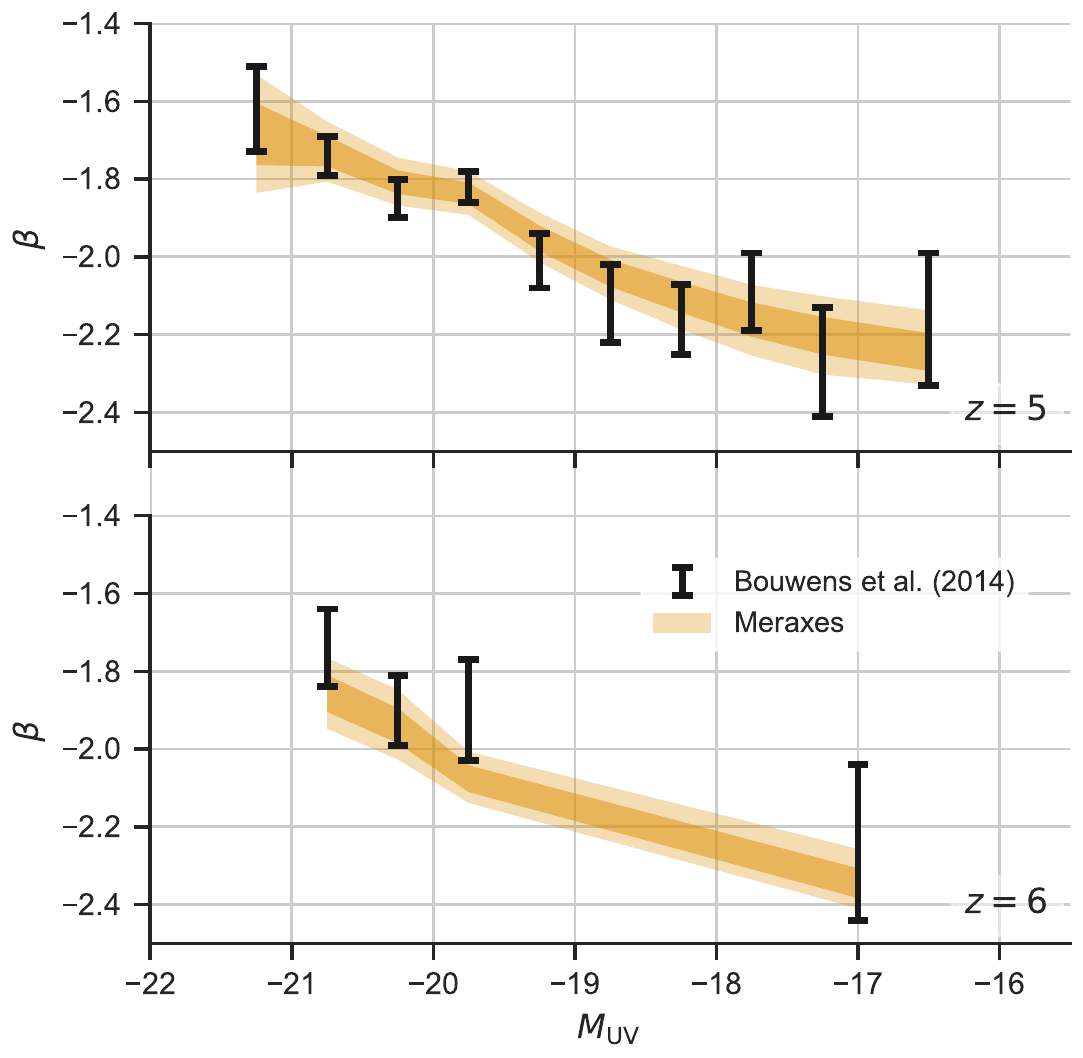}
  \vspace{-5mm}
  \caption{%
    The best-fit galaxy $\beta \textrm{-} M_{\rm UV}$ relation (shaded regions) obtained by fitting against the observed evolution of the UVLF and $\beta \textrm{-} M_{\rm UV}$ relation alone (error bars).
    Dark and light shaded regions indicate the 68 and 95\% confidence intervals, respectively.
    See \S\ref{sub:constraints_on_galform} for more details.
  }\label{fig:cmr-pred}
\end{figure}
We now explore the constraining power of the UVLF and $\beta - M_{\rm UV}$ relation on the galaxy formation parameters of our model.
In \citet{Mutch:2016}, we demonstrated that reionization feedback only plays a minor role in shaping the evolution of the observable ${z \gtrsim 5}$ galaxy population.
We therefore ignore our reionization constraints and turn off reionization feedback for this section, simplifying our interpretation and speeding up our calculations.

\begin{figure}
    \centering
    \includegraphics[width=1.0\columnwidth]{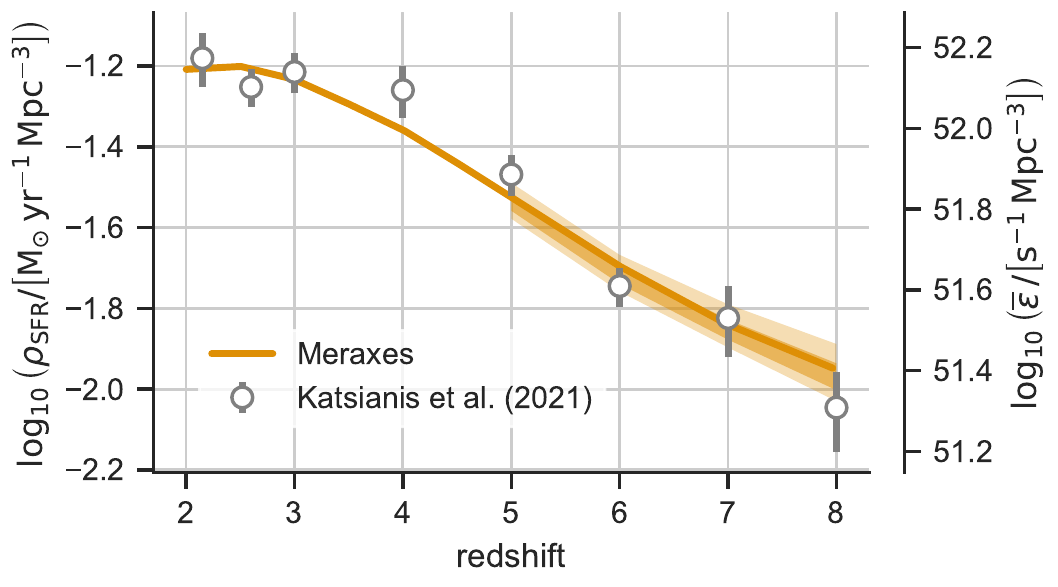}
    \vspace{-5mm}
\caption{%
 The predicted cosmic star formation rate density evolution obtained by fitting against the observed evolution of the UVLF and $\beta \textrm{--} M_{\rm UV}$ relation at $z{\geq}5$ (solid line).
 The right hand axes shows the corresponding volume averaged intrinsic ionizing emissivity.
 Dark and light shaded regions indicate the 68 and 95\% confidence intervals at $z{\geq}5$.
 Black data points show the observed star formation rate density from \citet{Katsianis2021} for comparison. Note that the model was not calibrated to reproduce the star formation rate density evolution specifically (see \S\ref{sub:low-z}).
}\label{fig:sfrd-pred}
\end{figure}

The model UVLFs are presented in Figure~\ref{fig:uvlf-pred}.
Shaded bands indicate the 68\% and 95\% confidence intervals of the model results, sampled from the underlying posterior distribution.
The model is able to replicate the observational data reasonably well, across all redshifts, however, the 68\% range of the model $\phi$ at fixed $M_{\rm UV}$ is noticeably smaller than that of the constraining observations.
This stems from a tension in matching all four redshifts simultaneously, especially around $-21 \lesssim M_{\rm UV} \lesssim -20$, where the observed $d\phi/dz$ varies non-monotonically.
When the model is constrained against a single redshift, the freedom afforded by the model and our choice of free parameters means that the magnitude of our uncertainties are similar to that of the constraining observations.
When fitting against multiple redshifts simultaneously, the volume of acceptable model parameters reduces such that we are no longer able to reproduce the full variance of a single redshift.
In Figure~\ref{fig:cmr-pred}, we show the corresponding $\beta - M_{\rm UV}$ relation.
Here, the uncertainty in the model results is more similar to that of the constraining observations at $z{=}5$. We also show in Figure~\ref{fig:sfrd-pred} the corresponding prediction for the star formation rate density (and volume averaged intrinsic ionizing emissivity) evolution together with their 68\% and 95\% confidence intervals. At redshift 8, the predicted emissivity can vary by $\sim25$\%, decreasing to $\sim 10$\% by $z\lesssim5$.

\begin{table*}
  \begin{minipage}{\textwidth}
    \centering
 \caption{The maximum a-posteriori (MAP) for each galaxy formation parameter when constraining against the observed evolution of the UVLF and $\beta$-$M_{\rm UV}$ relation. The 1-D marginalised MAP, mean, and 68 and 95\% highest density intervals (HDI) are also presented for each parameter. Plots of the 1-D and 2-D marginalised posterior distributions are presented in Figure~\ref{fig:uvlf-posteriors}.}%
    \label{tab:galform-params}%
    \begin{tabular}{l|c|cccc}
\hline
{} &     &  \multicolumn{4}{c}{marginalised} \\
   &    MAP    &  mode          &   mean         &        68\% HDI     &         95\% HDI \\
\hline
$\log_{10}(\alpha_{\rm SF})$ &  -1.11 &             -1.18 &  -1.20 &    [-1.26, -1.12] &    [-1.35, -1.05] \\
$\log_{10}(\Sigma_{\rm n})$  &  -0.85 &             -1.10 &  -1.50 &    [-1.76, -0.61] &    [-2.75, -0.57] \\
$\log_{10}(\epsilon_0)$      &   0.65 &              1.47 &   2.43 &      [0.62, 2.96] &      [0.51, 4.88] \\
$\log_{10}(\eta_0)$          &   0.68 &              0.64 &   0.68 &      [0.60, 0.75] &      [0.51, 0.84] \\
$\tau_{\rm ISM}$             &  14.00 &              9.41 &  18.66 &     [0.17, 22.08] &     [0.01, 46.66] \\
$\tau_{\rm BC}$              & 289.77 &            282.73 & 671.60 &  [105.27, 801.06] &  [65.63, 1650.87] \\
$\gamma_{\rm GCD}$           &   1.52 &              1.35 &   1.39 &      [1.15, 1.53] &      [1.06, 1.83] \\
$n$                          &  -2.43 &             -2.55 &  -2.46 &    [-2.88, -2.09] &    [-3.17, -1.64] \\
$a$                          &   0.69 &              0.88 &   0.87 &      [0.72, 0.97] &      [0.64, 1.11] \\
\hline
\end{tabular}

  \end{minipage}
\end{table*}

The maximum a-posteriori (MAP) parameter values, along with their marginalised posterior modes and highest density intervals (HDIs), are presented in Table~\ref{tab:galform-params}.
The corresponding one and two-dimensional posteriors are also shown in Appendix~\ref{sec:appendix-posteriors}.
With the exception of the critical surface density for star formation ($\Sigma_n$) and supernova energy coupling efficiency ($\epsilon_0$), all of the parameters possess a well-defined peak in their 1D marginalised posteriors which is independent of the imposed priors.
There are a number of degeneracies between the galaxy formation and dust model parameters. For example, a higher critical density for star formation ($\Sigma_n$) leaves more cold gas available for efficient star formation in massive systems and hence requires more dust attenuation there (controlled by $\gamma_{\rm GCD}$) to match the bright end of the UVLF.\@
The similarity of the $\epsilon_0$ posterior distribution to its prior at all but the smallest values indicates that the model's supernova feedback energy coupling has a minimum viable efficiency, beyond which there is little change to the predicted UVLF. This is due to saturation effects. Once $\epsilon_0$ is large enough, all available cold gas (as determined by the mass loading factor, $\eta_0$, and available cold gas reservoirs in the galaxy) is ejected and increasing the value of $\epsilon_0$ has no effect.

\subsection{Low redshift predictions}
\label{sub:low-z}

\begin{figure}
   \centering
   \includegraphics[width=1.0\columnwidth]{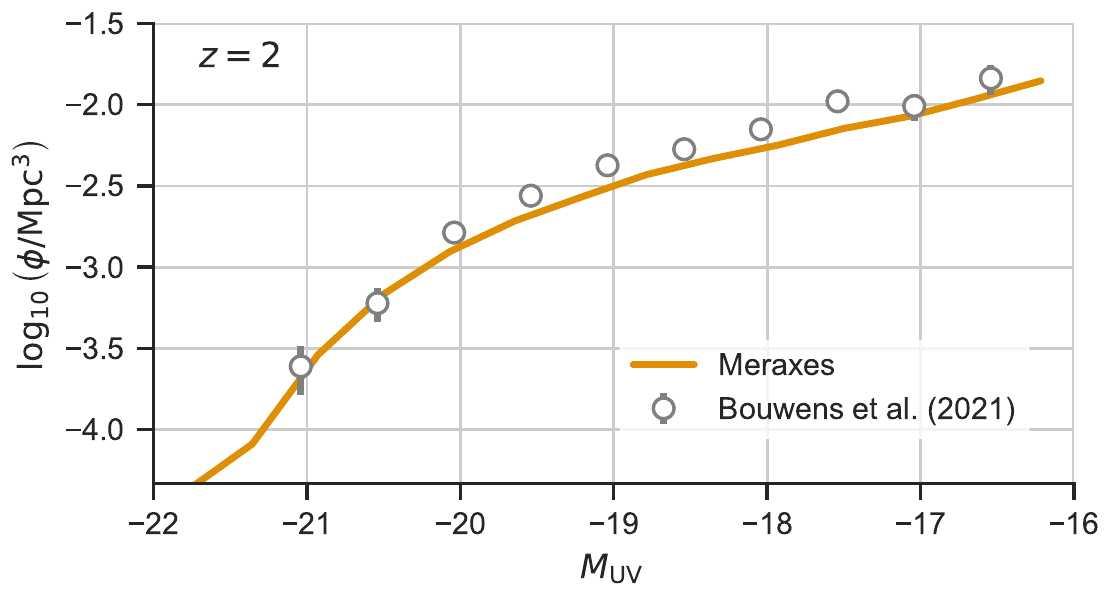}
 \vspace{-5mm}
 \caption{%
 The predicted $z{=}2$ UVLF found by constraining against the observed evolution of the UVLF and $\beta \textrm{--} M_{\rm UV}$ relation at $z{\geq}5$ (solid line). The observational results of \citet[]{Bouwens:2021} are shown for comparison (error bars). Note that the model was not calibrated to reproduce the UVLF at $z{<}5$ specifically (see \S\ref{sub:low-z}).
}\label{fig:uvlf-z2}
\end{figure}

For this work, we choose to constrain our galaxy formation model parameters against the $z{\geq}5$ observations alone.
Whilst \textsc{Meraxes} is capable of running all the way to $z{=}0$, there are numerous biases and complications associated with simultaneously constraining against high and low -redshift observations.
The abundance of available observations at low redshift, coupled with the typically smaller measurement uncertainties, means that our fit would prefer matching them, at the expense of matching potentially inconsistent high-redshift data.

Having said that, any model which completely fails to reproduce relevant low-redshift observations should be deemed invalid.
For this reason, we ran our model with the maximum a-posteriori galaxy formation parameters down to $z=1.8$ (the lowest redshift available from the Tiamat simulation) to ensure our predictions look sensible.
Figure~\ref{fig:sfrd-pred} shows the predicted evolution of the cosmic star formation rate density compared against the UV-derived observational results reported by \citet{Katsianis2021}.
Figure~\ref{fig:uvlf-z2} further shows the predicted $z=2$ UVLF compared against the observations of \citet{Bouwens:2021}.
In both cases there is a reasonable level of agreement between our best fit model and the observations, despite the fact that the model was not directly calibrated to achieve this.
We are therefore satisfied that our best-fit galaxy formation parameters are valid, at least for the purposes of reproducing the UV evolution of the galaxy population.

\section{Observational constraints on reionization}%
\label{sec:constraints}

Whilst the evolution of the observed UV luminosity function allows us to constrain the ionizing photon budget as a function of time, it does not provide any direct constraint on the fraction of these photons which escape their source galaxy and participate in reionization ($f_{\rm esc}$).
For this, we turn to the following three observations.

\subsection{Thompson scattering optical depth}%
\label{sub:taue-constraint}

The integrated optical depth due to Thompson scattering of cosmic microwave background photons by free electrons can be expressed as:
\begin{eqnarray}
  \begin{aligned}
    \tau_e = &\int_{z=0}^{\infty}\frac{c\,{\rm d}t}{{\rm d}z}{(1+z)}^3\sigma_{\rm T} \\
             &\times\left[Q^{\rm m}_{\rm HII}\left<n_{\rm H}\right> +
             (Q^{\rm m}_{\rm HeII} + 2Q^{\rm m}_{\rm HeIII})
             \left<n_{\rm He}\right>\right]\,{\rm d}z\ ,
  \end{aligned}
\end{eqnarray}
where ${\sigma_{\rm T}=6.652\times 10^{-25}\,{\rm cm^2}}$ is the Thomson scattering cross-section, $Q^{\rm m}_X$ is the mass-weighted global ionized fraction of species $X$, and ${\left<n_{\rm H}\right> = 1.88\times 10^{-7} (\Omega_{\rm b}h^2/0.022)\, {\rm cm^{-3}}}$ and ${\left<n_{\rm He}\right> = 0.148\times 10^{-7} (\Omega_{\rm b}h^2/0.022)\, {\rm cm^{-3}}}$ are the average comoving density of hydrogen and helium, respectively \citep{Wyithe:2003}.
In order to calculate $\tau_e$ from our model we follow the assumptions outlined in \citet{Mutch:2016}.
In particular, we assume that helium and hydrogen are singly ionized at the same rate (i.e. ${Q^{\rm m}_{\rm HeII}(z) = Q^{\rm m}_{\rm HII}(z)}$ and ${Q^{\rm m}_{\rm HeIII}({z>4})=0.0}$) until ${z=4}$, after which helium is assumed to be doubly ionized \citep[i.e. ${Q^{\rm m}_{\rm HeIII}({z\le 4})=1.0}$]{Kuhlen:2012}.

We utilise the \citet{Planck-Collaboration:2016} measurement of $\tau_e=0.058 \pm 0.012$ as our single data point.
As with the ionizing emissivity constraint above, we use use a standard $\chi^2$ likelihood and assume Normally distributed errors.
We further assume that the uncertainty on the model prediction is negligible and that there is no missing or underestimated systematic uncertainties associated with the observational data.

\subsection{Ionizing emissivity}%
\label{ssub:ionem-constraint}

The instantaneous ionizing emissivity (the number of meta-galactic ionizing photons per unit time, per Hydrogen atom\footnote{Often commonly expressed as the number of ionizing photons per unit time, per unit volume}) at the latter stages of reionization, can be inferred from Lyman-alpha forest observations \citep[e.g.][]{Bolton:2007, McQuinn:2011, Becker:2013}.
Following \citet{Mutch:2016}, we calculate this quantity from \textsc{Meraxes} using:
\begin{equation}
  \dot N_{\gamma} = \frac{\dot m_* N_{\gamma/m_\star}f_{\rm
  esc}}{f_{\rm b}M_{\rm tot}(1-0.75 Y_{\rm He})}\ ,
\label{eqn:dNiondt}
\end{equation}
where $\dot m_*$ is the total star formation rate in the full simulation volume, $N_\gamma/m_\star$ is the number of ionizing photons per stellar baryon (fixed at 6000 to be consistent with our assumed \citet{Kroupa:2001} IMF, ${f_b=\Omega_b/\Omega_m}$ is the universal baryon fraction, and $Y_{\rm He}$ is the Helium mass fraction.

In this work, we constrain against the emissivities calculated by \citet{DAloisio:2018}.
These were derived by combining the most recent measurements of the IGM opacity and gas temperature out to ${z\sim6}$ with a suite of high-resolution hydrodynamic simulations covering a broad range of IGM thermal histories.
For more details on both the methodology used to obtain the observed ionizing emissivity values and the treatment of the associated errors, we refer the reader to \citet{DAloisio:2018}. We utilise a standard $\chi^2$ likelihood assuming Normally distributed and uncorrelated errors (c.f. equation~\ref{eq:uvlf-lnL}).
We choose to restrict ourselves to $z>5$ in this work. While observations are available at $z<5$, and these ionizing emissivity data would place tighter constraints, the contribution from AGN can become non-negligible at $z \lesssim 4$ \citep[e.g.][]{Trebitsch:2021,FG:2020}. Furthermore, at $z=5$ the snapshot cadence of the Tiamat N-body simulation changes from constant in time to logarithmic in redshift, introducing complications in any analysis between $z\sim 4.5$ and 5.

\subsection{Global neutral fraction}%
\label{sub:xHI}

The volume averaged global neutral fraction can be probed observationally by a number of indirect methods.  These include Lyman-$\alpha$ source counts \citep[e.g.][]{McGreer:2015}, Lyman-$\alpha$ and Lyman-break galaxy clustering \citep[e.g.][]{Mason:2018,Hoag:2019}, and QSO damping-wing absorption features \citep[e.g.][]{Greig:2017a,Davies:2018,Greig:2019,Wang:2020}.
Due to the small available sample sizes and model-dependent nature of these probes, the associated uncertainties on individual measurements are typically large.
However, a direct measure of the volume averaged global neutral fraction, $\bar x_{\rm HI}$, can also be obtained from the fraction of pixels with zero flux in high-$z$ quasar spectra \citep{Mesinger:2010}.
The appearance of these `dark' pixels in the Lyman-$\alpha$ and/or $\beta$ forests indicates the presence of intervening neutral hydrogen, providing an upper limit to the cosmological neutral hydrogen fraction.

In this work, we utilise a number of different neutral fraction constraints assembled from the recent literature,
including both model dependent and independent sources \citep{McGreer:2015,Mason:2018,Hoag:2019,Greig:2017a,Davies:2018,Greig:2019,Wang:2020}.
With the exception of the model independent dark pixel constraints, we restrict ourselves to only utilising observational values with error bars, rather than upper or lower limits.
This is largely motivated by the difficulty in sensibly assigning uncertainties to such limits.
However, limited investigation has further suggested that the inclusion of these limits, with the most conservative assumption of a step function likelihood, leads to very little additional constraining power on the evolution of our model escape fraction in any case.

We use the 1-sigma uncertainties on each data point as presented and, following \citet{Greig:2017}, we model upper and lower limits as a one sided Gaussian with a likelihood of 1 below (/above) the observed value.
The model predictions are obtained by calculating the fraction of cells in our reionization grid with a neutral fraction less than $10^{-4}$ at the central redshift of each data point.

\section{Results}%
\label{sec:results}

In this section we use our best fit model (MAP) combined with a flexible halo mass dependent model for escape fraction to constrain the relation between escape fraction and galaxy properties, and to compute their contribution to the ionizing photon budget.  

\subsection{Constraints on escape fraction evolution}%
\label{sub:constraints_on_fesc}

Recent numerical studies have indicated that there is a likely correlation between halo mass and the escape fraction of ionizing photons \citep[e.g.][]{Paardekooper:2015,Kimm:2017,Yeh:2023}
This is thought to be a result of low mass halos having shallower potential wells, allowing supernova to more efficiently clear high density gas in the IGM and provide channels for ionizing photons to escape the galaxy.
In this section, we thus investigate the constraining power of the Thompson scattering optical depth, ionizing emissivity, and neutral fraction evolution on $f_{\rm esc}$ as a function of halo mass.

We fix all galaxy evolution parameters to their maximum a-posteriori (MAP) values found in Section~\ref{sub:UVLF}, where we constrained the model against the evolution of the UVLF.\@
We note that this removes any extra scatter in the recovered escape fraction introduced by uncertainties in the evolution of ionizing source population, simplifying our interpretation. However the observed star-formation rate density is constrained by the luminosity function to a $\sim 10\%$ level at $z\sim6$ (see Figure~\ref{fig:sfrd-pred}), which is more precise than the observed uncertainties in the ionization rate in the IGM. The 10\% uncertainty on the star-formation rate density (which is linearly related to the escape fraction with-respect-to the timing of reionization) is also smaller than the fractional constraints on the escape fraction derived below. We therefore find that the effect of simultaneously allowing our galaxy formation parameters to be free, in addition to the reionization parameters, would be minimal.  

\begin{figure}
  \centering
  \includegraphics[width=\columnwidth]{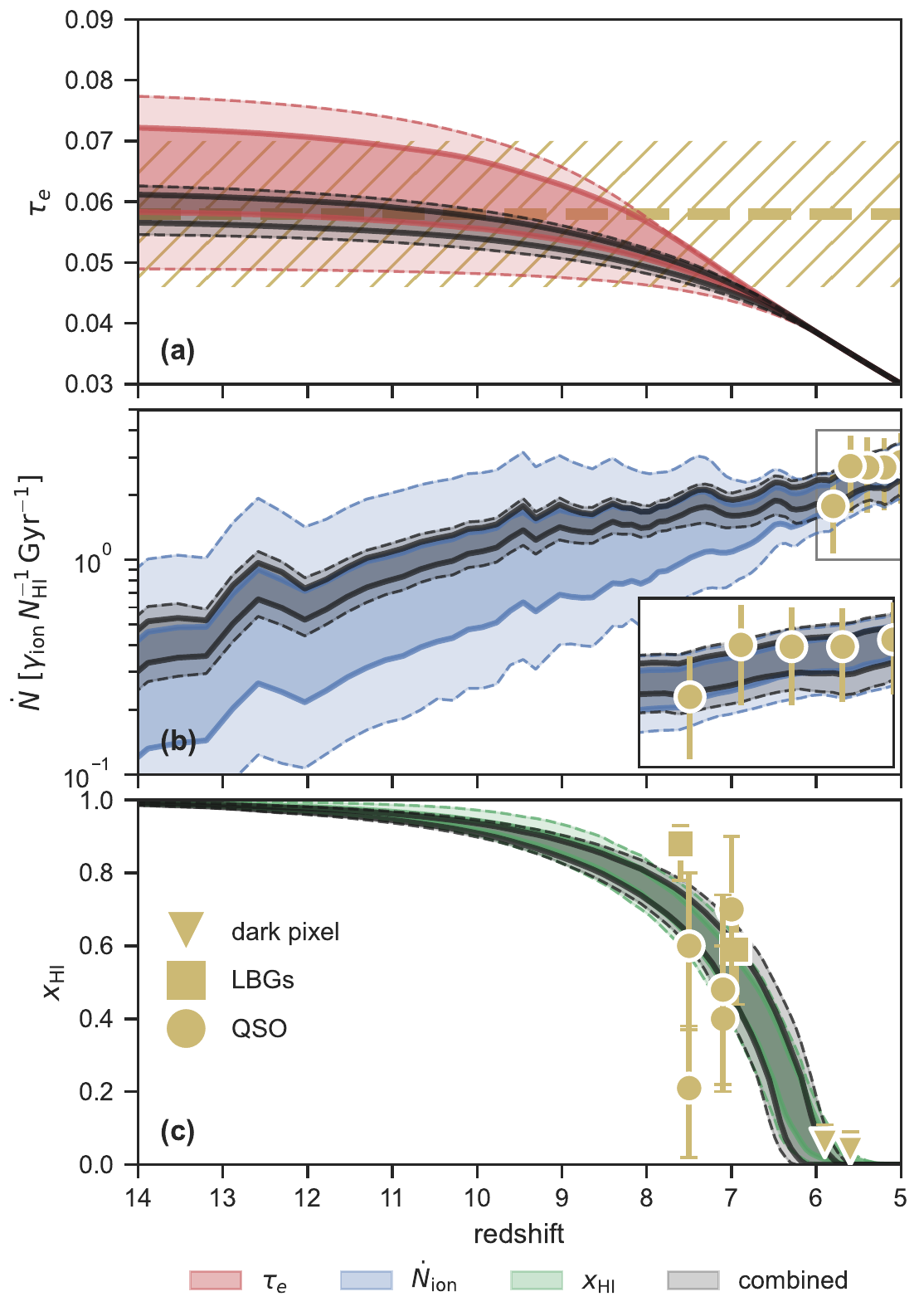}
\vspace{-5mm}
\caption{\label{fig:reion_only-constraints}%
    The reionization constraints used in this work (tan colour), along with the mass dependent escape fraction model constrained against each observation independently (red, blue and green shading) and simultaneously (black shading). Dark and light shading indicate the 68 and 95\% confidence intervals, respectively. The \textsc{Meraxes} galaxy formation parameters are fixed to their maximum a posteriori values as presented in table~\ref{tab:galform-params}. See \S\ref{sub:constraints_on_fesc} for discussion of these results.
    \textit{Panel (a):} The Thompson scattering optical depth (\S\ref{sub:taue-constraint}). 
    \textit{Panel (b):} The evolution of the total ionizing emissivity (\S\ref{ssub:ionem-constraint}). The inset panel shows a zoom-in of $z$=5-6.
    \textit{Panel (c):} The evolution of the volume weighted neutral fraction (\S\ref{sub:xHI}).
  }
\end{figure}

\begin{figure}
  \centering
  \includegraphics[width=0.9\columnwidth]{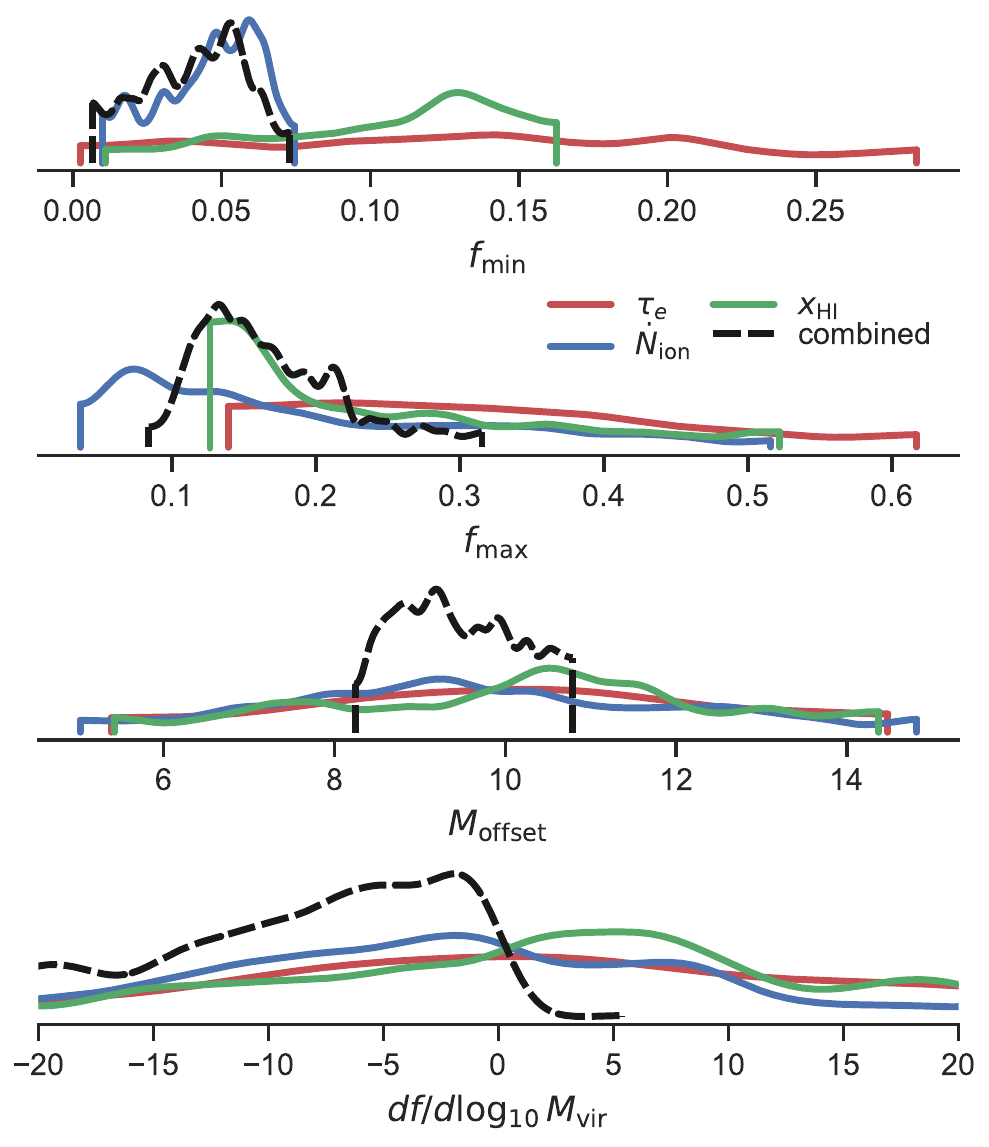}
\vspace{-2mm}
\caption{\label{fig:reion_only-split_posteriors} Kernel density estimate plots of the 1-D marginalised constraints on the escape fraction parameters (Equation~\ref{eqn:fesc_model}). The red, green and blue curves show the constraints from the CMB optical depth, neutral fraction and ionizing photon rate observations respectively. The black curves show the combined constraints.} 
\end{figure}

We choose a logistic functional form for the escape fraction:
\begin{equation}
  \label{eqn:fesc_model}%
  f_{\rm esc}(M_{\rm vir}) = f_{\rm min} + \frac{f_{\rm max} - f_{\rm min}}{1 + \exp\left(-\gamma_{\rm esc}\log_{10}\left(\frac{M_{\rm vir}}{M_{\rm offset}}\right)\right)}
\end{equation}
where
\begin{equation}
  \gamma_{\rm esc} = \frac{4 {df\over d\log_{10}M_{\rm vir}}}{\left(f_{\rm max} - f_{\rm min}\right)}
\end{equation}
This flexible, four parameter function allows the escape fraction to vary between $f_{\rm min}$ and $f_{\rm max}$ with a maximum rate of $df/\left(d\log_{10}M_{\rm vir}\right)$ midway between the two, at a mass of $M_{\rm offset}$.
We use non-flat priors for these parameters, both to encode our physical intuition, and to improve the efficiency of our sampling.
More details can be found in Appendix~\ref{sec:appendix-posteriors}, however, we have ensured that the precise choice of priors does not influence results except where explicitly discussed below.

The results of the model after constraining against the three observations are presented in Figures~\ref{fig:reion_only-constraints} and the 1-D marginalised posterior distributions of the escape fraction parameters are presented in Figure~\ref{fig:reion_only-split_posteriors}. Coloured lines and shaded regions show the results of fitting the escape fraction to match each observation independently, whilst black lines and shaded regions show the results of constraining against all three observational datasets simultaneously.

\paragraph*{Thompson scattering optical depth:}
The Thompson scattering optical depth is an integrated quantity spanning the full history of the IGM back to the surface of last scattering.
As such, it is sensitive to both the timing and effective mid-point of reionization.
From panel (a) of Figure~\ref{fig:reion_only-constraints} we see that the model is able to reproduce the \citet{Planck-Collaboration:2016a} observations (dashed line and hatched region) when using a galaxy growth history consistent with observations.

The kernel density estimate (KDE) of the 1-D marginalised constraints on the escape fraction parameters are shown by the red lines in Figure~\ref{fig:reion_only-split_posteriors}.
From here we can see that the optical depth provides very little constraining power on the escape fraction.
It merely demands that the escape fraction value be between $\sim 5 - 35\%$ across the full range of contributing halo masses present in the model ($10^8 \leq M_{\odot} \lesssim 10^{13}$).

Interestingly, our model disfavours $\tau_e \lesssim 0.52$, even when not requiring a simultaneous match to the observed ionizing emissivity and neutral fraction evolution (red shaded regions).
This suggests that achieving lower optical depths requires an escape fraction model which has a stronger scaling with redshift than is possible with halo mass alone.

More recent determinations of the optical depth from \citet{Planck:2020} suggest $\tau_e = 0.0544^{+0.007}_{-0.008}$. This is lower than the constraint used in this work (see Section~\ref{sub:taue-constraint}) and would require a more rapid reionization with a larger minimum $f_{\rm max}$ value than we see in Figure~\ref{fig:reion_only-split_posteriors}.
The result would be an increasing tension with our constraints on $\dot{N}_{\rm ion}$ and $x_{\rm HI}$, both of which prefer lower $f_{\rm max}$ values.

\paragraph*{Ionizing emissivity:}%
The total ionizing emissivity ($\dot N$) directly constrains the number of ionizing photons entering the IGM and thus, when coupled with a fixed UVLF evolution, places the strongest constraints on the escape fraction.
From the top two panels of Figure~\ref{fig:reion_only-split_posteriors} we see that ionizing emissivity prefers a mass independent escape fraction ($f_{\rm min} \approx f_{\rm max} \approx 7 \%$).
For escape fractions parametrisations which vary with mass, there is a preference for a large positive slope ($df/d\log_{10}M_{\rm vir}$) centred around $M_{\rm vir} \sim 10^9 - 10^{10}\, \mathrm{M_{\odot}}$.
This corresponds to a rapidly increasing escape fraction with decreasing halo mass in our parametrisation, resulting in a constraint at $M_{\rm vir} = 10^8\, \mathrm{M_{\odot}}$ of $f_{\rm esc} \sim 5 - 25 \%$.
This is driven by the observed flattening in the ionizing emissivity beyond $z \lesssim 6$, requiring both a rapid increase in the emissivity using early-forming, low mass haloes and a reduced contribution from high-mass haloes to keep the emissivity as flat as possible at low redshift (see blue shaded regions in panel (b) of Figure~\ref{fig:reion_only-constraints}).

\paragraph*{Neutral fraction evolution:}%
The range of allowed neutral fraction evolution models is shown by the green shaded regions in panel (c) of Figure~\ref{fig:reion_only-constraints}.
The dark pixel fraction upper limits of \citet{McGreer:2015} ensure that reionization is all but complete by $z{\sim}5.5$.
Combined with the higher redshift constraints, particularly that of \citet{Hoag:2019}, the escape fraction parameters are driven towards producing a late and rapid reionization scenario, requiring a minimum escape fraction of $\sim 12 \%$ across all halo masses (68\% confidence).
The KDEs of the 1-D marginalised parameter distributions in Figure~\ref{fig:reion_only-split_posteriors} again indicate that these observations are consistent with a mass independent escape fraction.
However, there is a weak preference for a positive slope ($df/d\log_{10}M_{\rm vir}$) and an increase in escape fraction with increasing halo mass.
This is contrary to our physical intuition and results from the desire to have as rapid evolution in the neutral fraction as possible, whilst maintaining a late start by keeping the contribution from early-forming, low mass haloes as small as allowed. 

\paragraph*{Combined reionization constraints:}%

\begin{figure}
  \centering
  \includegraphics[width=\columnwidth]{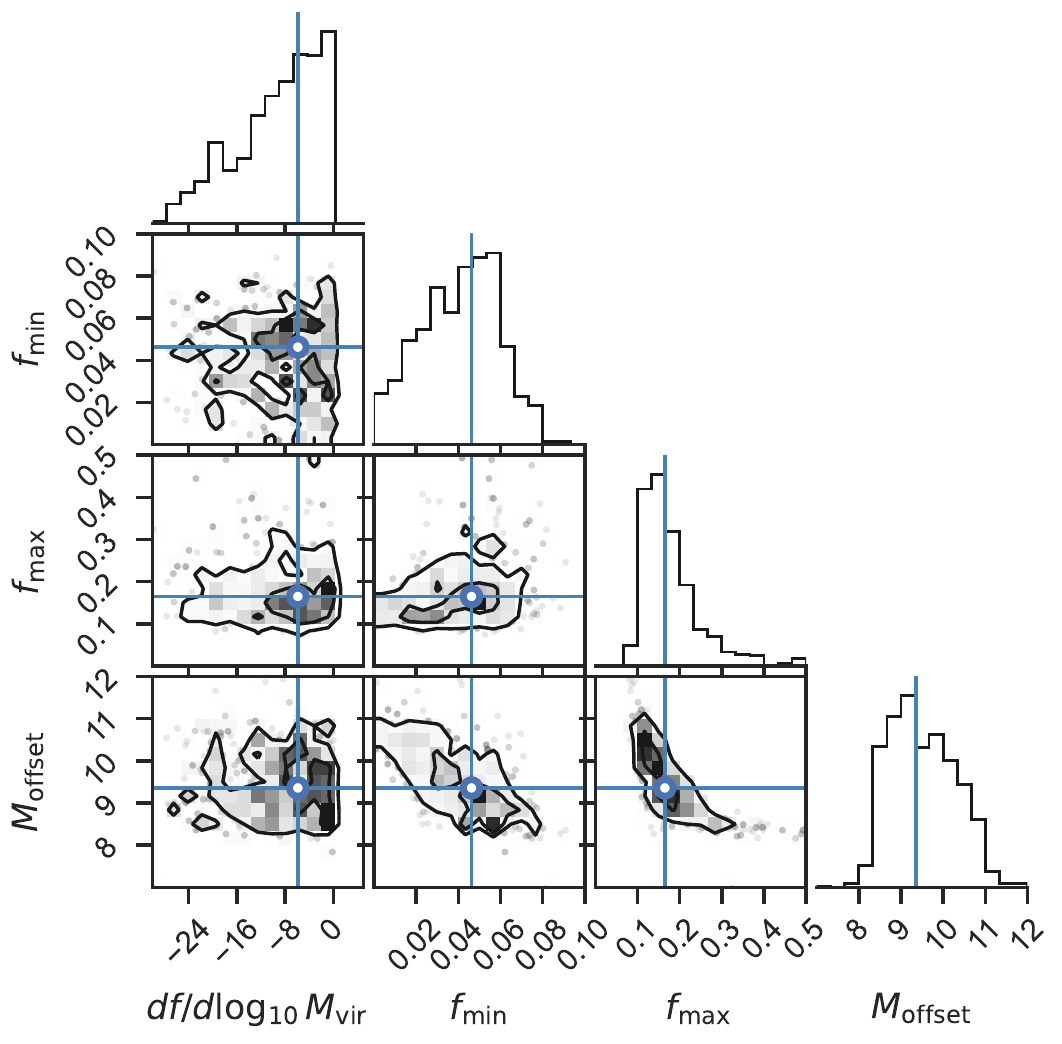}
\vspace{-5mm}
\caption{%
    The 1 and 2-D marginalised posterior distributions for our escape fraction model parameters when fitting to all three observational constraints simultaneously. Blue cross-hairs indicate the maximum a-posteriori parameter set. The 2-D contours represent 1 and 2-$\sigma$ probability density regions.
  }\label{fig:reion_only-posteriors-corner}
\end{figure}

\begin{figure}
  \centering
  \includegraphics[width=\columnwidth]{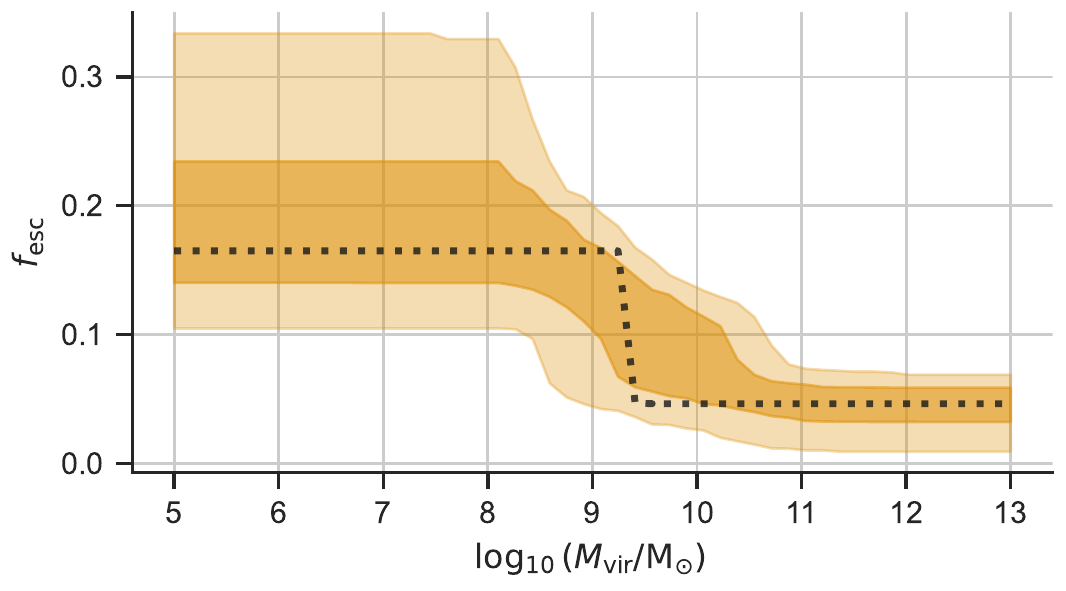}
\vspace{-5mm}
\caption{%
    The predicted escape fraction evolution when constrained against the Thompson scattering optical depth, ionizing emissivity, and neutral fraction evolution.
    The galaxy formation parameters are fixed to the  maximum a-posteriori (MAP) values found in Section~\ref{sub:UVLF}.
    Dark and light shaded regions indicate the 68 and 95\% confidence intervals, respectively. The dotted line shows the MAP escape fraction evolution.
    A general trend of increasing escape fraction with decreasing halo mass can be seen, in good agreement with both theory and other works.%
  }\label{fig:reion_only-posteriors}
\end{figure}

\begin{table*}
    \centering
    \caption{The maximum a-posteriori (MAP) for each escape fraction model parameter (Equation~\ref{eqn:fesc_model}) when constraining against the observed Thompson scattering optical depth, ionizing emissivity evolution, and global neutral fraction evolution. The 1-D marginalised mode, mean, and 68\% and 95\% highest density intervals (HDI) are also presented for each parameter. Plots of the 1 and 2-D marginalised posterior distributions are presented in Figures~\ref{fig:reion_only-split_posteriors} and \ref{fig:reion_only-posteriors-corner}, respectively.}%
    \label{tab:reiononly-params}%
    \begin{tabular}{l|c|cccc}
\hline
{} &    MAP &  \multicolumn{4}{c}{marginalised} \\
   &        &  mode          &   mean         &        68\% HDI     &         95\% HDI \\
\hline
$df/d\log_{10}M_{\rm vir}$ & -5.89 &             -2.63 &             -21.73 &        [-0.11,-15.74] &       [-0.11,-100.26] \\
$f_{\rm min}$              &  0.05 &              0.05 &               0.04 &           [0.03,0.06] &           [0.00,0.07] \\
$f_{\rm max}$              &  0.16 &              0.13 &               0.18 &           [0.10,0.20] &           [0.08,0.34] \\
$M_{\rm offset}$           &  9.36 &              9.18 &               9.52 &           [8.38,9.97] &          [8.33,10.94] \\
\hline
\end{tabular}
\end{table*}

The black shaded regions in Figure~\ref{fig:reion_only-constraints} and black dashed lines Figure~\ref{fig:reion_only-split_posteriors} show the result of calibrating the escape fraction model using all three reionization constraints together. For these combined constraints we also present the 2-D marginalised distributions for escape fraction model parameters in Figure~\ref{fig:reion_only-posteriors-corner} and the corresponding summary statistics in Table~\ref{tab:reiononly-params}.
From Figure~\ref{fig:reion_only-constraints} we can see that the model is able to successfully match all of the observational constraints simultaneously.
By including all constraints, we also greatly reduce the allowed range of model results, most notably those of the Thompson scattering optical depth (panel a) and ionizing emissivity (panel b).
Figures~\ref{fig:reion_only-split_posteriors} and \ref{fig:reion_only-posteriors-corner} show that a negative $df/d\log_{10}M_{\rm vir}$ (increasing escape fraction with decreasing halo mass) is strongly favoured.
The midpoint of the transition between $f_{\rm min}$ and $f_{\rm max}$ is also restricted to a much narrower range than was the case when constraining against any one observation alone.

In Figure~\ref{fig:reion_only-posteriors} we show the resulting constraint on the escape fraction as a function of halo mass.
At $M_{\rm vir} \gtrsim 10^{10.5}\, \mathrm{M_{\odot}}$ the escape fraction is $\sim 5\%$
This is driven largely by a low escape fraction in high mass halos being needed to keep the ionizing emissivity as flat as possible at $z = 5 - 6$.
The lack of ionizing photons entering the IGM from high mass haloes reduces the ability of the model to achieve a rapid reionization, producing optical depths $\gtrsim 0.062$ and a slower neutral fraction evolution (see panels (a) and (c) of Figure~\ref{fig:reion_only-constraints}).
At lower masses, the escape fraction rises to $\sim 15 - 30 \%$, required to start reionization early enough to match all three constraints.

The dotted line in Figure~\ref{fig:reion_only-posteriors} shows the maximum aposteriori escape fraction parameterisation (see Table~\ref{tab:reiononly-params} for the corresponding parameter values). It is tempting to draw conclusions about physical processes colluding to put some particular significance on halo masses of $M_{\rm vir}\approx 10^{9.35}\,{\rm M_\odot}$, where the sharp transition in $f_{\rm esc}$ occurs. However, the marginalised 68\% highest density interval for the $M_{\rm offset}$ parameter is broad ($10^{8.38} - 10^{9.97}\,{\rm M_\odot}$; Table~\ref{tab:reiononly-params}). From Figure~\ref{fig:reion_only-posteriors-corner}, we can also see that the position of the step is correlated with the maximum and minimum escape fraction parameters. Indeed, this correlation is what drives the appearance of a gradual evolution in 68 and 95\% confidence intervals of $f_{\rm esc}$ vs. $M_{\rm vir}$ in Figure~\ref{fig:reion_only-posteriors}.

\begin{figure}
  \begin{center}
    \includegraphics[width=\columnwidth]{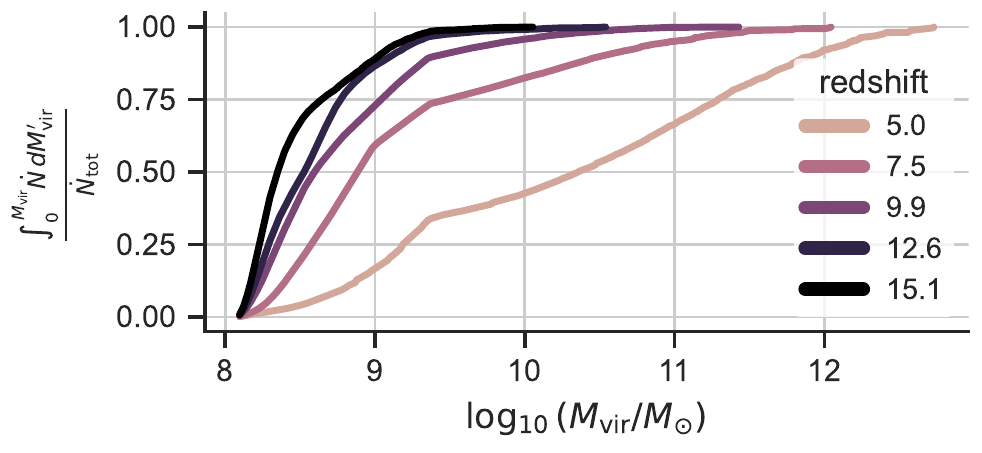}
  \end{center}
\vspace{-5mm}
\caption{The cumulative contribution to the instantaneous ionizing photon rate from halos below a given mass at different redshifts resulting from the best fit $f_{\rm esc}\left( M_{\rm vir}\right)$ model constrained to match all reionization datasets simultaneously.} 
  \label{fig:ndot-mvir-cumulative}
\end{figure}

Figure~\ref{fig:ndot-mvir-cumulative} shows the corresponding cumulative fractional contribution of halos to the instantaneous ionizing photon rate from halos below a given mass at different redshifts. As expected the typical halo mass producing ionizing photons increases during reionization. However, as noted above, the best fit $f_{\rm esc}\left( M_{\rm vir}\right)$ model favours an escape fraction that is dominated by low mass galaxies, while reionization also supresses star formation in low mass halos. Together these effects result in a very rapid transition in the halo mass above which 50\% of photons are produced from $M_{\rm vir}\gtrsim10^{8.5}$M$_\odot$ at $z\sim9$ to $M_{\rm vir}\gtrsim10^{10.5}$M$_\odot$ at $z\sim5$.

\subsection{Escape fraction evolution with galaxy properties} 
\label{ssub:fesc_with_galprops}

Previous studies of escape fraction evolution, both using simulations and observations, have found mean trends with galaxy properties such as stellar mass and colour \citep[e.g.][]{Pahl:2022,Begley:2022,Yeh:2023}.
In this section, we use the MAP $f_{\rm esc}\left(M_{\rm vir}\right)$ parameter values to investigate the predicted evolution of the escape fraction with stellar mass and star formation rate.

\begin{figure}
  \begin{center}
    \includegraphics[width=\columnwidth]{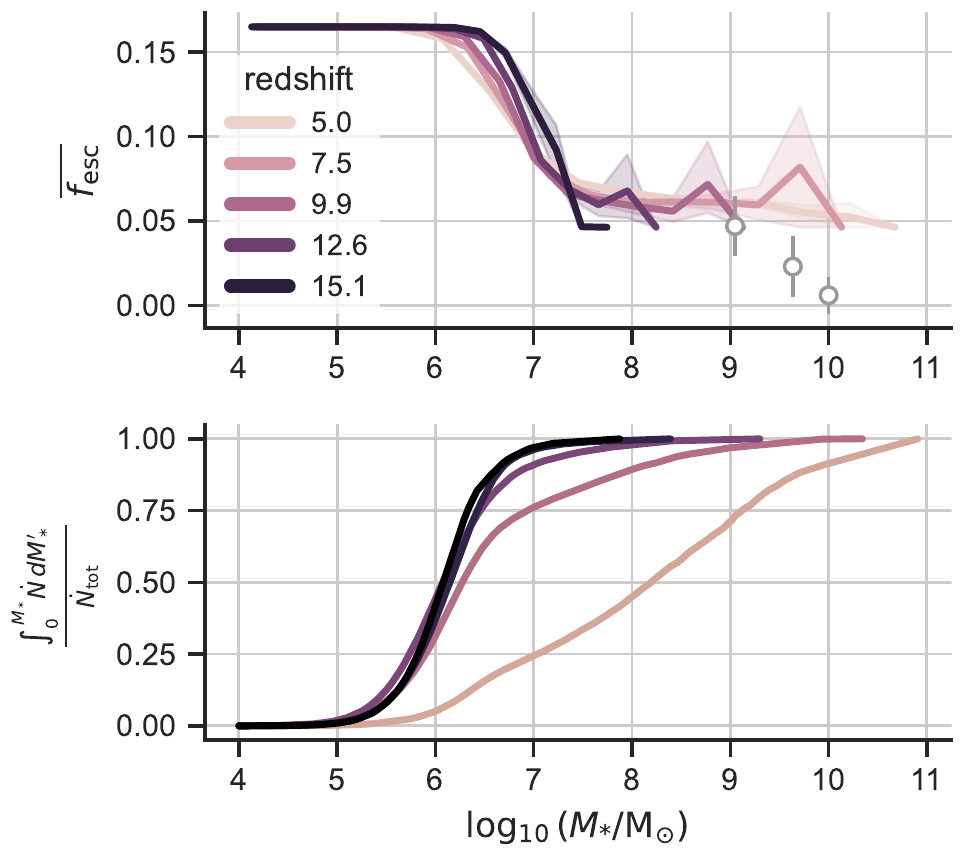}
  \end{center}
  \vspace{-5mm}
  \caption{Upper panel: The evolution of mean escape fraction vs. galaxy stellar mass resulting from the best fit $f_{\rm esc}\left( M_{\rm vir}\right)$ model constrained to match all reionization datasets simultaneously. Results are shown at 5 values of redshift.
  Shaded regions represent the 95\% confidence intervals of the mean, calculated from 1,000 bootstrapped samples.
  Data points are the observed $z{\sim}3$ results of \citet{Pahl:2022}. We highlight that these measurements were not included as constraints on our model. Lower panel: The corresponding cumulative contribution to the instantaneous ionizing photon rate from halos below a given stellar mass.}
  \label{fig:fesc-stellarmass}
\end{figure}

The upper panel of Figure~\ref{fig:fesc-stellarmass} shows the mean escape fraction as a function of galaxy stellar mass resulting from the $f_{\rm esc}\left( M_{\rm vir}\right)$ model (Equation~\ref{eqn:fesc_model}) with the MAP parameter values listed in Table~\ref{tab:reiononly-params}.
The shaded region indicates the 95\% confidence intervals on the mean, generated from 1,000 bootstrap samples. Since stellar mass correlates positively with halo mass in our model \citep{Mutch:2016}, the results are qualitatively similar to the dependance on halo mass.
The typical escape fraction decreases from its maximum of $\sim 20\%$ at $M_\star {\lesssim} 10^6\,{\rm M_\odot}$ to $\sim 7\%$ at $M_\star {\gtrsim} 10^8\,{\rm M_\odot}$.
This decrease begins at $M_\star\sim 10^{6.5}\,\mathrm{M_\odot}$ at $z \sim 15$ and shifts to smaller masses as reionization progresses due to a mild evolution of the typical stellar mass hosted by $M_{\rm offset} = 10^{9.25} \mathrm{M_\odot}$ haloes, the pivot point of our MAP $f_{\rm esc}\left(M_{\rm vir}\right)$ model.

The lower panel of Figure~\ref{fig:fesc-stellarmass} shows the corresponding cumulative fractional contribution of galaxies below a given stellar mass to the instantaneous ionizing photon rate. Since stellar mass correlates positively with halo mass we see a similar trend as Figure~\ref{fig:ndot-mvir-cumulative} where the stellar mass above which galaxies dominate the ionizing photon budget increases during reionization. The stellar mass above which 50\% of photons are produced is predicted to be close to a constant during the reionization era ($M_\star\sim10^{6}$M$_\odot$ at $z\gtrsim 7.5$), but then increases rapidly to $M_\star\sim10^{8.5}$M$_\odot$ at $z\lesssim6$.

\begin{figure}
  \begin{center}
    \includegraphics[width=\columnwidth]{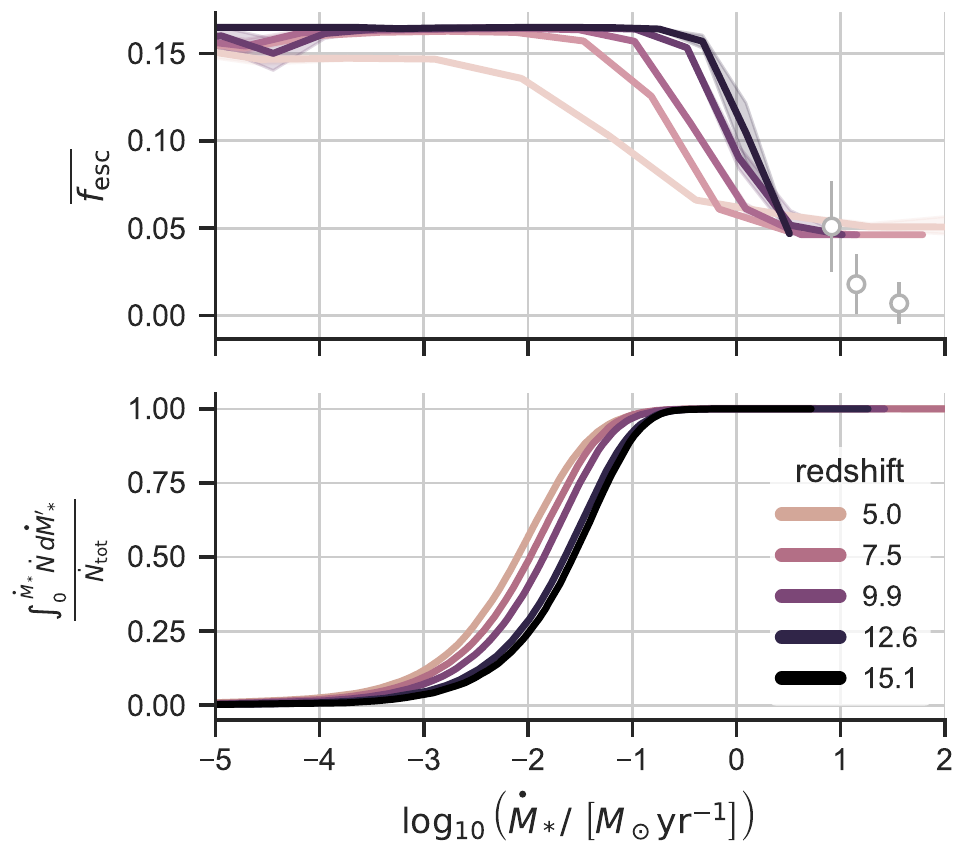}
  \end{center}
 \vspace{-5mm}
 \caption{Upper panel: The evolution of mean escape fraction vs. instantaneous star formation rate resulting from the best fit $f_{\rm esc}\left( M_{\rm vir}\right)$ model constrained to match all observational datasets simultaneously. Results are shown at 5 values of redshift.
  Shaded regions (only visible for $z{=}15$) represent the 95\% confidence intervals of the mean, calculated from 1,000 bootstrapped samples.
  Data points are the observed $z{\sim}3$ results of \citet{Pahl:2022}. We highlight that these measurements were not included as constraints on our model. Lower panel: The corresponding cumulative contribution to the instantaneous ionizing photon rate from halos below a given stellar mass.}
  \label{fig:fesc-sfr}
\end{figure}

Figure~\ref{fig:fesc-sfr} shows how the mean escape fraction instead varies by star formation rate for our MAP model.
We see a similar evolutionary trend as was found with stellar mass.
The escape fraction has a more extended evolution with star formation rate as redshift decreases. The lower panel of Figure~\ref{fig:fesc-sfr} illustrates that galaxies with star-formation rates between $\dot{M}_\star\sim10^{-1.5}$M$_\odot$yr$^{-1}$ and $\dot{M}_\star\sim10^{-2.5}$M$_\odot$yr$^{-1}$ contribute the dominant sources of ionizing photons throughout reionization.


\section{Discussion}%
\label{sec:discussion}

Before concluding we compare our population based findings with other recent theoretical results and observational studies.

Simulations typically show values of escape fraction that vary between a few and a few 10s of percent \citep[e.g.][]{Paardekooper:2015,Rosdahl:2022}, which is in agreement with our results and with prior empirical comparisons of galaxy luminosity functions and Ly-alpha forest constraints \citep[e.g.][]{Bolton:2007,Wyithe:2010}. Hydrodynamic radiative transfer simulations further indicate that escape fractions decrease with halo mass for halos with $M_{\rm vir} \lesssim 10^9\, \mathrm{M}_\odot$ \citep[e.g.][]{Kimm:2017,Lewis:2020}. Most recently \citet{Yeh:2023} have computed the escape fractions of reionizing galaxies within the \textsc{Thesan} project. They find that low-mass galaxies having $M_\star\lesssim10^7$M$_\odot$ are the main drivers of reionization above $z\gtrsim7$, while higher-mass galaxies with $M_\star\gtrsim10^8$M$_\odot$ dominate the escaped ionizing photon budget at lower redshifts. When calculating the dependence on mean escape fraction as a function of halo mass, \citet{Yeh:2023} report a value of a few percent in larger halos with a sharp rise towards halo masses below $M_{\rm vir}\sim10^9$M$_\odot$ \citep[see also][]{Kostyuk:2023}. Inference studies by \citet{Park:2019} also consider similar observational constraints to those used in this work (the UV luminosity function, CMB optical depth and dark pixels). Although their UV ionizing escape fraction has a simpler, power-law relation with the host halos mass, they also find $f_{\rm esc}$ around 5--30 per cent for low-mass haloes, with no significant preference on the power-law index. All of these results are in qualitative agreement with our findings, especially when the different simulation techniques and mode of analysis are considered. 

In addition to theoretical results we can compare our results in light of observational estimates of the escape fraction. Using stacked spectra of Lyman break galaxies at $z\sim3$ \citet{Pahl:2021} \citep[following][]{Steidel:2018} find an average value for the escape fraction of $0.06 \pm 0.01$. Subsequently \citet{Pahl:2022} find an anti-correlation between the escape fraction and stellar mass. In Figures~\ref{fig:fesc-stellarmass} and \ref{fig:fesc-sfr} we reproduce the measurements of escape fraction from \citet{Pahl:2022} for comparison with our model constraints (we stress that these measurements were not included as constraints on our model). The masses of these galaxies are at the upper end of those thought to be responsible for reionization ($M_\star\gtrsim10^8$M$_\odot$). However, our results overlap at the lowest stellar masses and star formation rates present in \citet{Pahl:2021}, as well as with many of the simulation studies referenced above. Whilst the correlations at high masses and star formation rates appear different, we note that we have constrained our escape fraction to follow a redshift-independent, halo mass dependent form. We have very few galaxies in these high stellar mass / star formation rate regimes due to our limited simulation volume and therefore they provide almost no contribution to our fits.

Finally, we note that, whilst our N-body simulation, Tiamat, provides a good balance between mass resolution and simulated volume, it does not fully resolve all haloes down to the redshift-dependent atomic cooling mass threshold. However, we miss less than 3 per cent of stellar mass at $z \approx 6$ and less than 35 per cent at $z \lesssim 12$, therefore we do not expect this to significantly alter our conclusions \citep[see Section~A1 of][]{Mutch:2016}.

\section{Conclusions}
\label{sec:conclusions}

We have used the \textsc{Meraxes} semi-analytic model to infer the mean ionizing photon escape fraction and its dependence on galaxy properties through joint modelling of the observed high redshift UV galaxy luminosity function and the available constraints on the reionization history with the \textsc{Meraxes} semi-analytic model. \textsc{Meraxes} couples galaxy formation to a semi-numerical description of reionization, allowing the exploration of reionization and the growth of the galaxy population both temporally and spatially.

In this paper we present the application of nested sampling to \textsc{Meraxes}, allowing statistical exploration of the underlying free parameter space, including identification of the best fit model as well as statistical confidence intervals. Using a flexible model that relates  the escape fraction to halo mass, we use the joint constraints of the UV luminosity function, CMB optical depth, and the Ly$\alpha$ forest to evaluate the allowed range of escape fraction values and halo mass dependence. Our results show that available constraints require an escape fraction of $(18\pm5)\%$ for galaxies within halos of $M\lesssim10^{9}$M$_\odot$ and $(5\pm2)\%$ for larger mass halos. 

Because \textsc{Meraxes} provides the star-formation rates and stellar masses for each dark matter halo at each simulation snapshot, we can use these results to estimate the dependence of escape fraction on stellar mass and redshift. We find that there is a transition from an escape fraction of $\sim18\%$ to a smaller value of $\sim5\%$ which occurs at stellar masses of $M_\star\sim10^7$M$_\odot$. We find that this transition value is quite insensitive to redshift. Early in reionization it corresponds to star formation rates of $\dot{M}_\star\sim10^{-1}$M$_\odot$yr$^{-1}$, and decreases towards $\dot{M}_\star\sim10^{-2}$M$_\odot$yr$^{-1}$ by the end of reionization at $z\sim5$.

When considering the contribution of galaxies to the ionizing photon budget as a function of redshift, we find that reionization is dominated by the smaller $M_\star\lesssim10^7$M$_\odot$ galaxies with high escape fractions at $z\gtrsim6$, and by the larger galaxies $M_\star\gtrsim10^7$M$_\odot$ with lower escape fractions at $z\lesssim6$. These stellar masses correspond to galaxies with star formation rates of between $\dot{M}_\star\sim10^{-1.5}$M$_\odot$yr$^{-1}$ and $\dot{M}_\star\sim10^{-2.5}$M$_\odot$yr$^{-1}$, which represent the dominant sources throughout reionization. In sum, these results agree with recent direct measurements of a $\sim5\%$ escape fraction from massive galaxies at the end of reionization, and support results from hydrodynamic modelling and direct observation indicating that the escape of ionizing photons is dominated by low mass galaxies during reionization.

\section*{Acknowledgements}

This research was supported by the Australian Research Council Centre of Excellence for All Sky Astrophysics in 3 Dimensions (ASTRO 3D), through project number CE170100013.
Parts of this work were performed on the OzSTAR national facility at Swinburne University of Technology.
This work was supported by software support resources awarded under the Astronomy Data and Computing Services (ADACS) Merit Allocation Program. ADACS is funded from the Astronomy National Collaborative Research Infrastructure Strategy (NCRIS) allocation provided by the Australian Government and managed by Astronomy Australia Limited (AAL).
This research relies heavily, and with great thanks, on the Python open source community, in particular \textsc{numpy} \citep{Harris:2020}, \textsc{scipy} \citep{Virtanen:2020}, \textsc{Cython} \citep{Behnel:2011}, \textsc{matplotlib} \citep{Hunter:2007}, \textsc{h5py}, \textsc{ultranest} \citep{Buchner:2021}, \textsc{seaborn} \citep{Waskom:2021}, \textsc{mpi4py} \citep{Dalcin:2005}, \textsc{jupyter} \citep{Granger:2021}, \textsc{jinja2}, \textsc{click}, \textsc{pandas} \citep{pandas}, and \textsc{xarray} \citep{Hoyer:2017}.

\section*{Data Availability}

The data underlying this article will be shared on reasonable request to the corresponding author.




\bibliographystyle{mnras}
\bibliography{mhysa.bib}



\appendix

\begin{table*} \
    \centering
    \caption{Description of the priors employed for all free parameters.}%
    \label{TabA1}
    \begin{tabular}{ll}
      Parameter & Prior \\
      \hline
      \multicolumn{2}{c}{\textit{Galaxy formation}} \\
      $\log_{10}\left(\alpha_{\rm SF}\right)$ & $f(x) = \textrm{Truncated Normal}\left(x, x_{\rm min}=-3.0, x_{\rm max}=0.0, \mu=-1.0, \sigma=1.05\right)$ \\
      $\log_{10}\left(\Sigma_{\rm n}\right)$ & $f(x) = \textrm{Truncated Normal}\left(x,  x_{\rm min}=-3.0, x_{\rm max}=0.0, \mu=-1.0, \sigma=1.05\right)$ \\
      $\log_{10}\left(\epsilon_0\right)$ & $f(x) = \textrm{Normal}\left(x, \mu=1.0, \sigma=2.0\right)$ \\
      $\log_{10}\left(\eta_0\right)$ & $f(x) = \textrm{Normal}\left(x, \mu=1.0, \sigma=2.0\right)$ \\
      \hline
      \multicolumn{2}{c}{\textit{Dust model}} \\
      $\tau_{\rm ISM}$ & $f(x) = \textrm{Beta}\left(x/65, \alpha=1.0, \beta=1.1\right)/65$ \\
      $\tau_{\rm BC}$ & $f(x) = \textrm{Beta}\left(x/5000, \alpha=1.0, \beta=1.8\right)/5000$ \\
      $\gamma$ & $f(x) = \textrm{Beta}\left(x/15, \alpha=1.0, \beta=1.35\right)/15$ \\
      $n$ & $f(x) = \textrm{Beta}\left(x/15, \alpha=1.0, \beta=1.35\right)/15$ \\
      $a$ & $f(x) = \textrm{Beta}\left(x/10, \alpha=1.0, \beta=1.5\right)/10$ \\
      \hline
      \multicolumn{2}{c}{\textit{Escape fraction model}} \\
      $f_{\rm min}$ & $f(x) = \textrm{Beta}\left(x, \alpha=1.0, \beta=3.0\right)$ \\
      $f_{\rm max}$ & $f(x) = \textrm{Beta}\left(x, \alpha=1.5, \beta=4.2\right)$ \\
      $M_{\rm offset}$ & $f(x) = \textrm{Beta}\left(x/20, \alpha=1.3, \beta=1.3\right)/20$ \\
      $df/d\log_{10}M_{\rm vir}$ & $f(x) = \textrm{Student-t}(x/10, \nu=1)/10$ \\
    \end{tabular}
\end{table*}

\section{Prior and posterior parameter distributions}%
\label{sec:appendix-posteriors}

\begin{figure*}
  \begin{minipage}{\textwidth}
  \centering
  \includegraphics[width=1.0\textwidth]{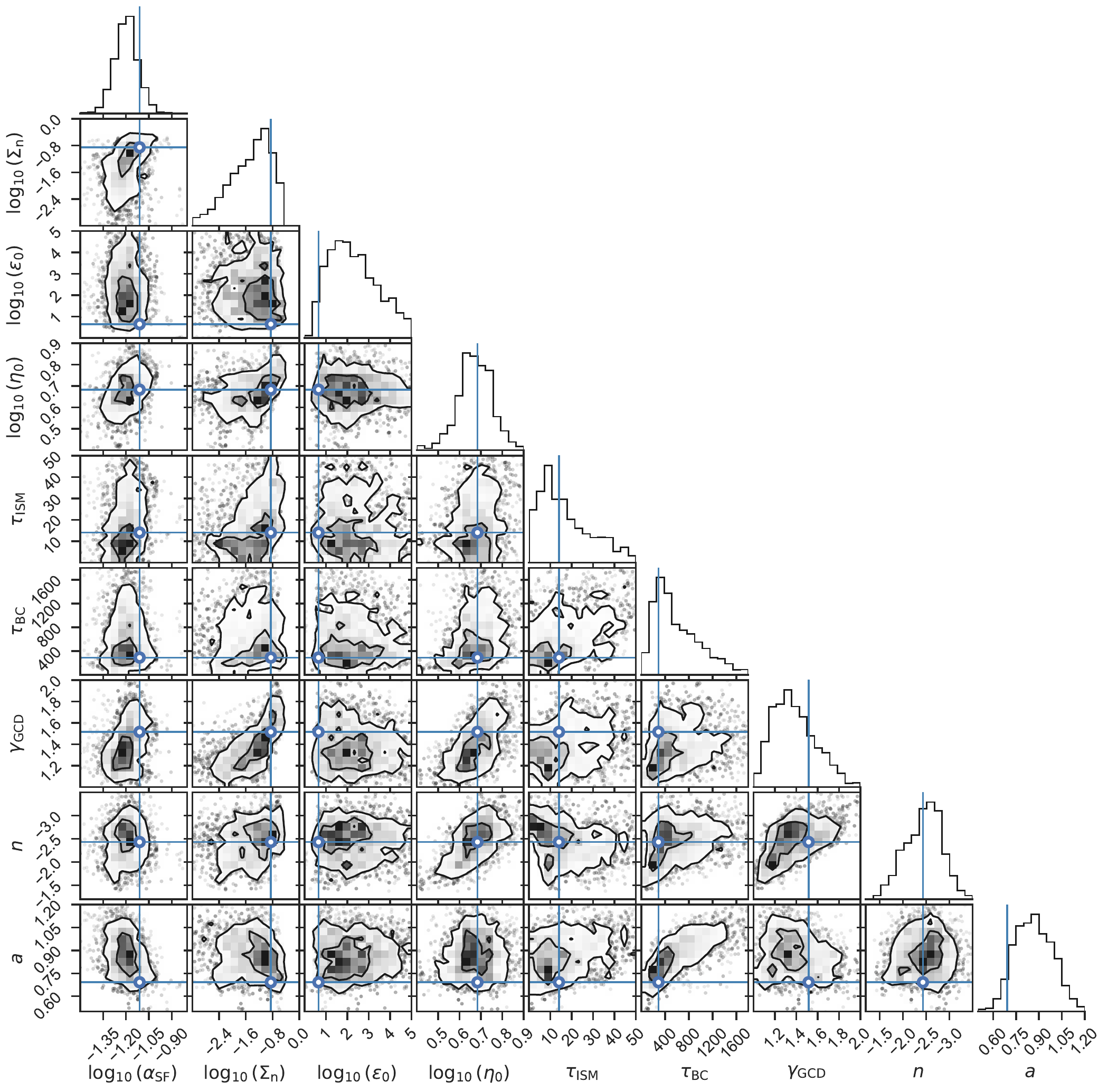}
   \vspace{-5mm}
   \caption{%
    The 1 and 2-D marginalised posterior distributions for our galaxy formation model parameters when constraining against the observed evolution of the UVLF and $\beta$-$M_{\rm UV}$ relation. Blue cross-hairs indicate the maximum a-posteriori parameter set. The 2-D contours represent 1 and 2-$\sigma$ probability density regions.
  }\label{fig:uvlf-posteriors}
  \end{minipage}
\end{figure*}

In this Appendix we present the 2-D marginalised probability distributions (Figure~\ref{fig:uvlf-posteriors}) for the fits of the \textsc{Meraxes} model to the galaxy UV luminosity function and $\beta-M_{\rm UV}$ relation discussed in Section~\ref{sub:constraints_on_galform}. It is worth noting that \citet{Qiu:2019} also used Meraxes with the same free galaxy formation and dust parameters as we do, but recovered different posterior distributions for some parameters (see their figure~A3). This can largely be attributed to the different redshift range utilised for their constraints ($z=7 - 4$). The inclusion of lower redshift data, in particular, provides significant extra constraining power. The remaining differences are due to our different choice of priors which can have an impact on poorly constrained parameters.

\begin{figure}
  \centering
  \includegraphics[width=\columnwidth]{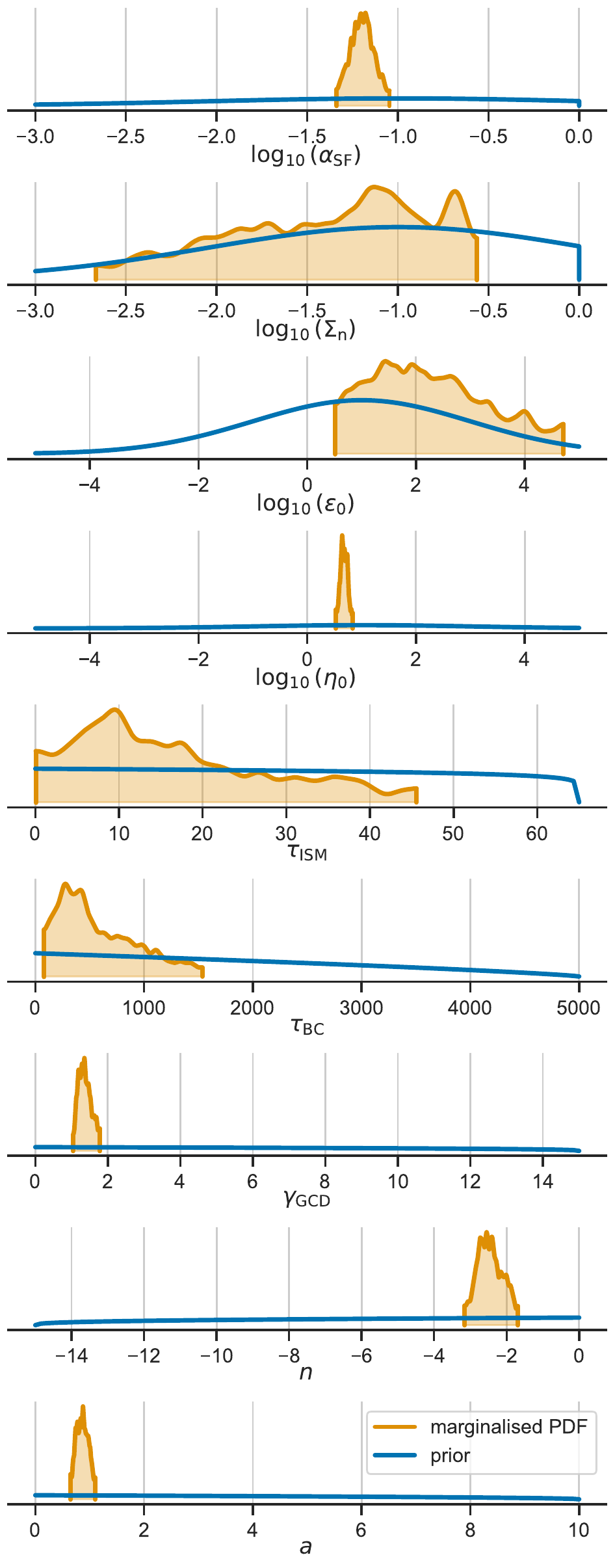}
    \vspace{-5mm}
  \caption{%
    Kernel density estimates of the marginalised PDFs for each galaxy formation parameter (orange; see Section~\ref{sub:constraints_on_galform}) compared with the corresponding prior distributions (blue lines; see Tables~\ref{tab:galform-params} \& \ref{TabA1}). A similarity between posterior distributions and priors indicates a lack of constraining power (e.g. $\log_{10}\left(\epsilon_0\right)$).
  }\label{fig:fesc_model_priors}
\end{figure}

\begin{figure}
  \centering
  \vspace{5mm}
  \includegraphics[width=\columnwidth]{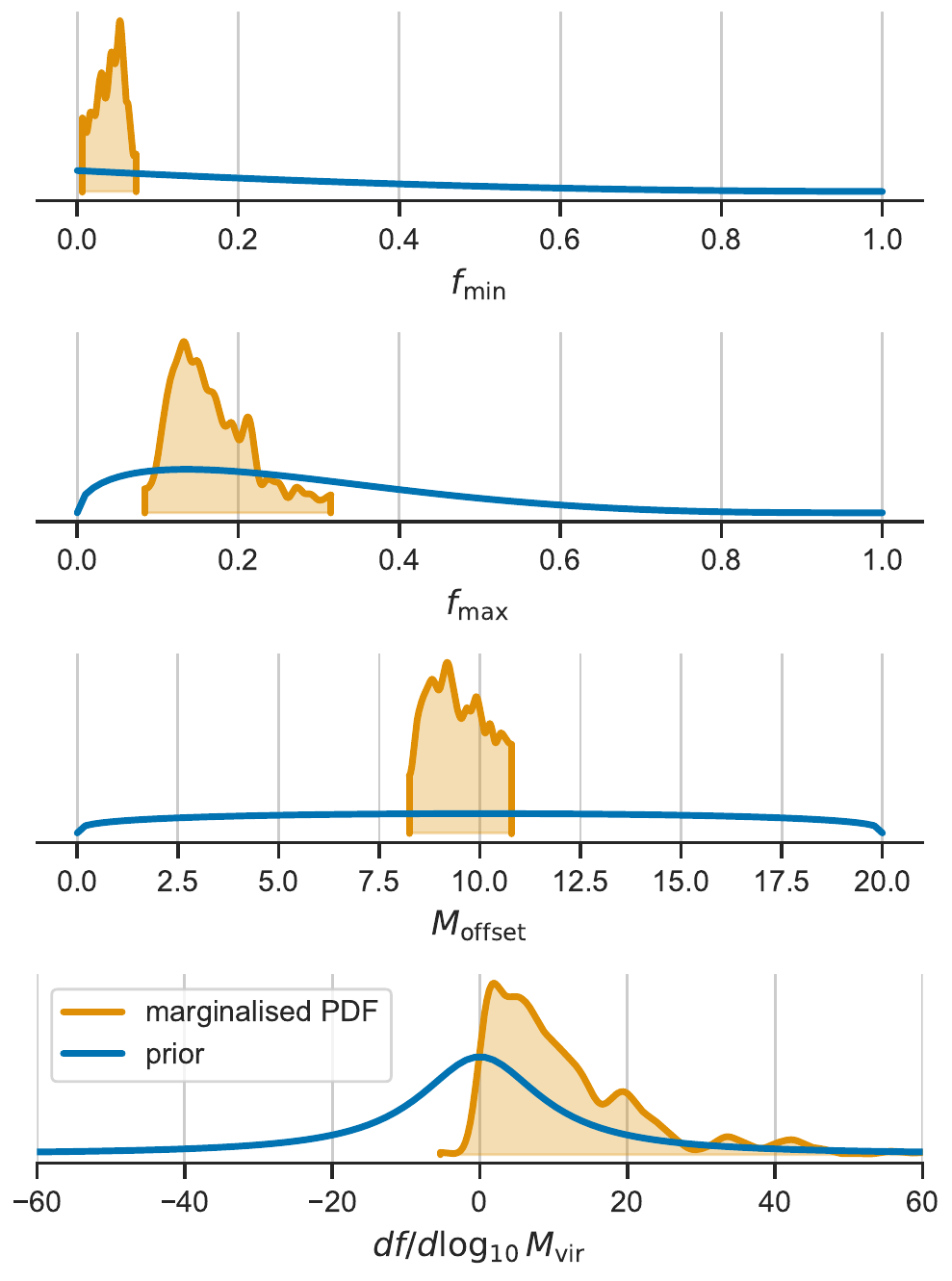}
   \vspace{-5mm}
   \caption{%
       Kernel density estimates of the marginalised PDFs for each $f_{\rm esc}$ model parameter when fitting against the combined reionization constraints (orange; see Section~\ref{sub:constraints_on_fesc}). These are compared with the corresponding prior distributions (blue lines; see Tables~\ref{tab:reiononly-params} \& \ref{TabA1}). A similarity between posterior distributions and priors would indicate a lack of constraining power.
  }\label{fig:fesc_model_priors-corner}
\end{figure}

We also provide detail of the prior probability distributions employed for all free parameters in this work. These encode the range in allowed values (e.g.\ efficiency parameters must be in the range 0--1). In some cases they also weakly encode physical intuition and the authors experience from past studies. We have ensured that the precise choice of priors makes no significant qualitative or quantitative difference to our findings except where explicitly noted in the text.  In appendix Table~\ref{TabA1} we present a tabulated description of the priors, and show these graphically in Figures~\ref{fig:fesc_model_priors} and \ref{fig:fesc_model_priors-corner}.

Note that we do not use uniform priors as commonly employed in the galaxy formation literature. Our reasoning is that flat priors give extra, undue, weight to parameter values which we know from experience and physical intuition to be unlikely. When using such priors, one also must choose bounds for the parameter values, which in itself often imposes prior knowledge or intuition anyway. The use of weakly informative priors, such as the ones we employ, additionally provides numerous other statistical benefits \citep[e.g.][]{Lemoine:2019}.  

\bsp{}\label{lastpage}

\end{document}